\documentclass[11pt,oneside]{article}

\usepackage[english]{babel}
\usepackage{tgpagella}
\usepackage{parallel}
\usepackage{setspace}

\usepackage{graphicx}
\usepackage[toc]{appendix}

\usepackage[dvipsnames]{xcolor} 
\usepackage{ulem}  
 
\usepackage{setspace}
\usepackage{caption}

\usepackage{amsfonts}
\usepackage{amssymb}
\usepackage{amsmath}
\usepackage{mathabx}
\usepackage{latexsym}
\usepackage{amsthm}
\usepackage{textcomp}
\usepackage{xcolor}
\usepackage{color}
\usepackage[dvipsnames]{xcolor} 
\usepackage{diagbox} 

\usepackage{hyperref}
\hypersetup{
    colorlinks, 
    linkcolor={red!50!black},  
    citecolor={blue!50!black},  
    urlcolor={blue!80!black}   
}

\usepackage{pstricks}
\usepackage{epigraph}
\usepackage{picture}
\usepackage[longnamesfirst]{natbib}

\def\reference#1{\href{#1}{Cliquer ici pour voir une r\'ef\'erence.}} 

\usepackage{tabularx}
\usepackage{thmtools}
\usepackage{thm-restate}
\usepackage{mathrsfs}

\definecolor{aquamarine}{rgb}{0, 0.60, 0.54}

\definecolor{brickred}{rgb}{0.8, 0.25, 0.33}

\newtheoremstyle{mytheoremstyle} 
    {\topsep}                    
    {\topsep}                    
    {\itshape}                   
    {}                           
    {\bf\scshape}                   
    {.}                          
    {.5em}                       
    {}  

\usepackage{sectsty} 
\sectionfont{\normalfont\scshape\centering}
\subsectionfont{\centering}
\providecommand{\U}[1]{\protect \rule{.1in}{.1in}}
{\normalfont\itshape}

\theoremstyle{mytheoremstyle}
\newtheorem{theorem}{Theorem} 
\newtheorem*{theorem*}{Theorem}

\newtheorem{lemma}{Lemma}
\newtheorem{corollary}{Corollary} 
\newtheorem*{corollary*}{Corollary} 
\newtheoremstyle{mydefinitionstyle} 
    {\topsep}                    
    {\topsep}                    
    {}                   
    {}                           
    {\bf\scshape}                   
    {.}                          
    {.5em}                       
    {}  
\theoremstyle{mydefinitionstyle}
\newtheorem{definition}{Definition} 
\newtheorem{example}{Example} 
\newtheorem{counterexample}{Counterexample}

\newtheorem*{question*}{Question}

\newtheorem{remark}{Remark}

\def\es{\varnothing}

\author{\textsc{Davide Carpentiere\thanks{University of Catania, Catania, Italy. davide.carpentiere@phd.unict.it}, Alfio Giarlotta\thanks{University of Catania, Catania, Italy. alfio.giarlotta@unict.it}, Angelo Enrico Petralia\thanks{University of Catania, Catania, Italy. angelo.petralia@unict.it}, Ester Sudano\thanks{Queen Mary University of London, London, United Kingdom, and University of Palermo, Palermo, Italy. e.sudano@qmul.ac.uk. Corresponding author.}}}

\title{\textbf{Separable joint choices
}\thanks{Angelo Petralia acknowledges the support of "Ministero del Ministero dell'Istruzione, dell'Universit\`a e della Ricerca (MIUR), PE9 GRINS "Spoke 8", project \textit{Growing, Resilient, INclusive, and Sustainable}, CUP E63C22002120006.}}

\newtheoremstyle{blueconj}{}{}{}{}{\bfseries\color{blue}}{.}{.5em}{}

\theoremstyle{blueconj}

\newcommand{\X}{\mathscr{X}}

\begin{document}
	
	\maketitle	
	
	\begin{abstract}
		\noindent 
		We introduce a novel choice dataset, called joint choice, in which options and menus are multidimensional. 
		In this general setting, we define a notion of choice separability, which requires that selections from some dimensions are never affected by those performed on the remaining dimensions.  
		This generalizes the classical definition of separability for discrete preference relations and utility functions, to encompass a class of choice behaviors that may lack a preference or utility representation.
		We thoroughly investigate the stability of separability across dimensions, and then suggest effective tests to check whether a joint choice is separable.
		Upon defining rationalizable joint choices as those explained by the maximization of a relation of revealed preference, we examine the interplay between the notions of rationalizability and separability.
		Finally, we show that the rationalizability of a separable joint choice can be tested by verifying the rationalizability of some derived joint choices over fewer dimensions.
	\end{abstract}
	




\section*{Introduction} \label{SECT:intro}

Choice typically happens within a space of multidimensional alternatives.
In this settings, it is relevant to study interdependencies across decisions at each dimension. 
A suitable notion of choice separability enables us to effectively deal with this situation. 
In this first work on the topic, we transpose and generalize the analogous concepts introduced for utility functions and preference relations.
Crucially, while classical separability is defined for utility representations that 
correspond to a class of rational choice behaviors, our framework directly defines separability in terms of observable choices. 
This approach disentangles separability from rationality, allowing us to analyze it as a standalone property.

To explicitly account for multidimensionality, we propose a novel dataset, called \textit{joint choice}. 
This novel setting represents selections from finite multidimensional menus, that is, vectors with finitely many components, each of which is a finite set of available alternatives associated to some dimension $q\in Q$.
Then \textit{separability} of a joint choice with respect to a subset $S \subseteq Q$ ($S$-\textit{separability}) requires that picks over dimensions that belong to $S\subseteq Q$ do not depend on alternatives offered in the dimensions belonging to $-S$, and separability of the whole dataset accounts for $S$-separability for any possible $S$. 
The proposed setting appears to be natural in several instances in which choices that involve multiple dimensions are of interest, such as modelling consumer's purchases of bundles of goods, choices over intertemporal streams of consumption or multi-step budgeting.
In such situations --- as well as for additional possible applications --- testing for separability can identify complementarities inherent in the objects of choice. 
Furthermore, joint choices can represent sequences of individual decisions, allocations of resources to a group of individuals, and choices by multiple members of the same social group. 
We show that separability in this cases can serve as a discriminant to classify deterministic models of choice and allocations, and can be tested as a proxy for independence across decisions at each dimension. 

We prove that $S$-separability does not necessarily imply $T$-separability, when $T$ is a superset or subset of $S$.
However, the joint satisfaction of $S$-separability and $T$-separability yields both $(S \cup T)$-separability and, under an additional property of the choice domain, $(S \cap T)$-separability.
Moreover, we show that the separability of a joint choice can be inferred from the separability of joint choices over fewer dimensions, obtained by considering the observed picks associated to suitable subsets of $Q$. 
In light of these findings, we suggest effective tests to refute separability, independent of any rationality assumptions.  

We also introduce a notion of \textit{rationalizability} for joint choices, drawing upon the classical revealed preference framework employed for choice correspondences. 
We prove that the maximization of $S$-separable discrete preferences over multidimensional alternatives generates $S$-separable joint choices.
On the other hand, $S$-separable rationalizable joint choices whose domain satisfies a given richness property reveal $S$-separable preferences.
Similarly, additively separable utility functions, as defined by \cite{Strotz1957,Strotz1959} and \cite{Gorman1959}, yield separable choices, but the converse is not always true.
Finally, we show that the rationalizability of a separable joint choice is characterized by the rationalizability of joint choices over fewer dimensions, obtained from the original dataset.

\subsubsection*{Relation with literature}

This work extends to the setting of joint choices the well-known notions of separability defined for preferences and utilities.  
In this respect, \cite{Strotz1957,Strotz1959} and \cite{Gorman1959}, and also \cite{BarberaSonnenscheinZhou1991}, \cite{BarberaMassoNeme1997}, and \cite{Aliprantis1997} investigate separable continuous preferences and representing utility functions, whose induced ranking over some dimensions does not depend on the remaining dimensions. 
A separable utility function yields a specific behavior: within a subset of dimensions $S$, the optimal allocation of expenditure for each dimension preserves relative proportions in response to changes in income or prices pertaining to other dimensions.
More recently, \cite{BradleyHodgeKilgour2005} analyze the structure of separable discrete preferences, showing that these objects do not satisfy some properties, such as stability under some set operations and additivity, which, as pointed out by \cite{Gorman1968}, hold for the classical continuous case. 
While our notion of separability is independent of rationality, we find that the maximization of separable preferences and utility functions yields separable joint choices. 

Separability of choice behavior has been addressed  only very recently. 
Specifically, \cite{ChambersMasatliogluTuransick2024} define separable two-dimensional joint choice functions, in which the selection from a menu belonging to one dimension is not affected by the other dimension.
By defining and analyzing separability of joint choices, we extend their approach to more than two dimensions.  
It is worth mentioning that \cite{ChambersMasatliogluTuransick2024} also propose a stochastic generalization of their approach --- later characterized by \cite{KashaevPlavalaAguiar2024} --- which allows to study correlation between the choice of two agents. 
However, in this paper we focus our analysis on deterministic joint choices. 


An additional strand of research in which our work may offer some advancements is revealed preference theory.
In the traditional approach, originated from the work of \cite{Samuelson1938} on choice functions, and later extended by \cite{Ricther1966} to choice correspondences, rationalizability is defined as the possibility of generating the observed choice through the maximization of a preference elicited from the agent's selections over consumption bundles. 
In this classical setting, the multidimensionality of the menus is usually implicit.
On the contrary, in our contribution we make this hypothesis explicit, and study rationalizable joint choices, which are obtained by maximizing the preference revealed from the agent's choice over multidimensional alternatives.
In this extended framework, if separability holds, then rationalizability is equivalent to the rationalizability of special families of joint choices on fewer dimensions retrieved from the original dataset. 
\smallskip

The paper is organized as follows.
In Section~\ref{SECT:separability} we define separable joint choices. 
In Section~\ref{SECT:separability_behaves_well} we assess the stability of separability across dimensions, give a useful characterization, and show that separability discriminates a class of choice models and allocations. 
In Section~\ref{SECT:testing_separability} we derive some results that reduce the computations needed to test separability on data.
In Section~\ref{SECT:rational_vs_separable} we introduce rationalizable joint choices; upon comparing rationalizability and separability of joint choices, we show that the latter extends separability of discrete preferences and utility functions, and characterize the former.
Section~\ref{SECT:conclusions} collects some concluding remarks.




\section{Choice separability} \label{SECT:separability}


The object of our analysis is choice behavior within spaces of multidimensional alternatives. 
In this section, first we introduce joint choices, which explicitly comprise all separate dimensions that can be identified in a decision problem.   
Then we give a formal definition of separability within this multidimensional setting, and provide some enlightening examples. 


\subsection*{Joint choices} \label{SUBSECT:joint_choice}

For simplicity, first we suppose there are only two dimensions, that is, $Q =\{1,2\}$. 
We denote $X_1$, $X_2$ the sets of all alternatives associated to each dimension. 
We call a \textit{menu} an ordered pair $A_{1,2} = (A_1, A_2)$, such that $A_1 \subseteq X_1$ and $A_2 \subseteq X_2$, from which a selection is made. 
Possible menus are collected in $\mathfrak{M}_{1,2}$.
Selected items are in turn pairs $(a_1, a_2)$ such that $a_1 \in A_1$ and $a_2 \in A_2$.

\begin{definition} \label{DEF:joint_choice_for_2}
	A \textit{joint choice on} $\mathfrak{M}_{1,2} \subseteq 2^{X_1}\times 2^{X_2}$ is a map $c \colon \mathfrak{M}_{1,2} \to 2^{X_1 \times X_2}$ such that
	$$
	\es \neq c(A_1,A_2) \subseteq A_1 \times A_2
	$$
	for all $(A_1,A_2) \in \mathfrak{M}_{1,2}$.\footnote{\label{FOOTNOTE:simple_notation}Unless confusion may arise, to enhance readability, we shall often simplify notation, and omit round and curly brackets as well separators. For instance, above we have used $\mathfrak{M}_{1,2}$ in place of $\mathfrak{M}_{\{1,2\}}$, and $c(A_1,A_2)$ in place of $c((A_1,A_2))$. Similarly, a choice of the type $c(\{a,b\},\{d,e,f\})$ will be simplified into $c(ab,def)$, etc.}  
	If $\mathfrak{M}_{1,2} = 2^{X_1}\times 2^{X_2}$, $c$ is a \textit{complete} joint choice.
\end{definition}
%

\begin{example}[\it Consumption choices]
	Suppose each $q \in \{1,2\}$ represents a category. 
	Specifically, $1$ is the index for snacks, with all possible options in set $X_1$, and $2$ is the index for beverages, where $X_2$ is the set of all options. 
	The Cartesian product $X_1 \times X_2$ is the set of all possible bundles that contain a product per category. 
	Given a menu including available products for each category, $A_{1,2} = (A_1, A_2) = (\{\text{\it scones, croissants, cantucci}\}, \{\text{\it tea,  coffee}\}) \in \mathfrak{M_{1,2}}$, a joint choice selects bundles among those available, $c(A_1,A_2) = \{(\text{\it scones, tea}), (\text{\it cantucci, coffee})\}$.
\end{example}	

It is worth noting that the definition of joint choices presents a departure from the classical paradigm of choice in economics, given by the difference between domain and codomain. 
Nevertheless, this feature appears natural for our setting and useful for our analysis. 
The domain $\mathfrak{M}_{1,2} \subseteq 2^{X_1}\times 2^{X_2}$ ensures that selections are not limited to prearranged combinations of items. 
The codomain $2^{X_1 \times X_2}$ mimics the possibility to select specific bundles, rather than all possible combinations of some items.
We only require that, for any menu and any dimension, the associated pick be nonempty.

\begin{remark}[\it {Extending the domain of complete joint choices}] 
	A complete joint choice $c \colon 2^{X_1} \times 2^{X_2} \to 2^{X_1 \times X_2}$ can be extended to a choice correspondence $C \colon 2^{X_1 \times X_2} \to 2^{X_1 \times X_2}$ by defining it on $2^{X_1 \times X_2} \setminus \left( 2^{X_1} \times 2^{X_2}\right)$.
	Note that the image $c (A_1,A_2)$ of a menu $(A_1,A_2) \in 2^{X_1} \times 2^{X_2}$ is a subset of $A_1 \times A_2$ that need not be representable as $B_1 \times B_2$ for some $B_1 \subseteq A_1$ and $B_2 \subseteq A_2$. 
	For instance, if $c$ is a joint choice correspondence on $2^{X_1 \times X_2 }=2^{ \{a,b\} \times \{x,y\}}$ such that $c(X_1,X_2)=\{(a,x),(b,y)\}$, then $c(X_1,X_2)$ cannot be represented as a Cartesian product of menus. 
\end{remark}

To extend Definition~\ref{DEF:joint_choice_for_2} to an arbitrary finite set $Q$ of dimensions, we start by establishing some basic terminology.
For clarity, we itemize notation:
\begin{itemize}
	\item $Q$ is a nonempty finite set of indices, representing the \textit{dimensions} involved in a decision problem; for the sake of clarity, henceforth we set $Q = \{1,2,\ldots, n\}$ for some $n \geqslant 2\,$;
	\item $X_q$ is the nonempty finite set of all the \textit{one-dimensional alternatives} associated to the dimension $q \in Q\,$;
	\item $2^{X_q}$ is the family of \textit{one-dimensional menus} (i.e., nonempty subsets of $X_q$), where $q \in Q\,$;
	\item $\prod_{q \in Q} X_q$ is the set of all \textit{(multidimensional) alternatives};
	\item $x_Q= (x_q)_{q \in Q}$ is the generic (multidimensional) alternative in $\prod_{q \in Q} X_q\,$;
	\item $A_Q =(A_q)_{q \in Q} \in \prod_{q \in Q} 2^{X_q}$ is a \textit{(multidimensional) menu}, with $\es\neq A_q \subseteq X_q\,$;
	\item $\mathfrak{M}_Q\subseteq  \prod_{q \in Q} 2^{X_q}$ is an arbitrary nonempty family of (observable) finite menus.
\end{itemize}
\smallskip

\begin{definition} \label{DEF:joint_choice_general}
	A \textit{joint choice} on $\mathfrak{M}_{Q}\subseteq \prod_{q\in Q} 2^{X_q}$ is a map  such that\vspace{-0,1cm}
	$$
	c \colon 	\mathfrak{M}_{Q} \to 2^{\prod_{q \in Q} X_q} \;\; \text{such that} \;\;\es \neq c(A_Q) \subseteq \prod_{q \in Q}A_q \;\; \text{for all}\;\; A_Q \in \mathfrak{M}_{Q}.\vspace{-0,2cm}
	$$
Moreover, $c$ is \textit{complete} if $\mathfrak{M}_{Q}=\prod_{q\in Q} 2^{X_q}.$
\end{definition}

%
   
Let us clarify from the outset that the ordering of indices in the product plays no role in our analysis. 
As a consequence, we may assume without loss of generality that this ordering is (consistently) fixed on both the set $Q$ and each subset $S$ of $Q$.

For instance, suppose $Q=\{1,2,3\}$ and let
	\begin{itemize}
		\item $X_1=\{a,b\}=ab$,\footnote{As said in Footnote~\ref{FOOTNOTE:simple_notation}, we simplify notation, omitting curly brackets and separators whenever unambiguous.}, $X_2=\{p,q,r\}=pqr$, $X_3=\{w,x,y,z\}=wxyz$, 
		\item $\mathfrak{M}_{1,2,3}=\{(ab,pq,wxy),(a,pqr, yz),(b,qr,wz)\}$, i.e., there are only three feasible menus.
	\end{itemize}
	Define a joint choice $c\colon \mathfrak{M}_{1,2,3}\to 2^{X_{1}\times X_{2}\times X_{3}}$ by
	\begin{align*}
		c(ab,pq,wxy) &= \{(a,p,w),(a,q,x),(a,p,y)\}, \\
		c(a,pqr,yz) &= \{(a,r,y),(a,q,y)\}, \\
    	c(b,qr,wz) &= \{(b,q,z)\}.  
	\end{align*}
	Thus, the joint choice $c$ selects three (tridimensional) alternatives from the menu $(ab,pq,wxy)$, two alternatives from the menu $(a,pqr,yz)$, and only one alternative from the menu $(b,qr,wz)$.


\subsection*{Separability} \label{SUBSECT:separability}

Here we introduce the main notion of the paper, and prove a few preliminary results. 
We first establish some additional notation that helps us referring to relevant entries in menus and alternatives, whenever a subset $S$ of dimensions is fixed. 
For any nonempty set $S \subseteq Q$, set:
\begin{itemize}
	\item $-S = Q \setminus S$, the \textit{complement} of $S$;
	\item $\prod_{q \in S} X_q$, the set of all (multidimensional) alternatives having only dimensions in $S$;
	\item $x_S= (x_q)_{q \in S}$, the generic (multidimensional) alternative in $\prod_{q \in S} X_q$.
\end{itemize}
Furthermore, for any $\es \subsetneq S \subseteq T \subseteq Q$, the \textit{projection of $T$ onto $S$} is the map 
$$
\pi_S^T \colon \prod_{q \in T} X_q \to \prod_{q \in S} X_q\,,\;\; \text{defined by} \;\; \pi_S^T(x_T) = x_S \;\; \text{for all}\;\;x_T \in \prod_{q \in T} X_q\,.
$$ 
In particular, for $T = Q$ we write $\pi_S \colon \prod_{q \in Q} X_q \to \prod_{q \in S} X_q$ for the \textit{projection onto $S$}.\footnote{If $S = Q$, then $\pi_Q$ is the identity map on $X_Q$.}  

Given an alternative $x_T$, the projection of $T$ onto $S$ simply isolates the entries in $x_T$ that have indices in $S$, to obtain a smaller vector $x_S$.
Note that for any $x_Q$ such that $\pi_T(x_Q) = x_T$, we have $\pi_S^T(x_T) = \pi_S(\pi_T(x_Q)) = \ \pi_S(x_Q)$.
As usual, we denote by $\pi_S(A) \subseteq \prod_{q \in S} X_q$ the image of $A \subseteq \prod_{q \in Q} X_q$ under $\pi_S$.

To extend projections to set-functions, we use the map
$$
\pi_S \colon \prod_{q \in Q } 2^{X_q} \to \prod_{q \in S } 2^{X_q}\,,\;\; \text{defined by} \;\; \pi_S(A_Q) = A_S \;\; \text{for all}\;\; A_Q \in \prod_{q \in Q } 2^{X_q}
$$
where $A_S$ is the \textit{partial menu} obtained as the projection of the full menu $A_Q$ on the given dimensions in $S$, that is,\vspace{-0,1cm}
$$
A_S=\{x_S \in \prod_{q \in S} X_q \colon x_S = \pi_S(x_Q) \text{ for some } x_Q \in A_Q\}.
$$
%

\begin{example}[\it Consumption choices, continued]
	Consider the alternative $x_{1, 2}= (\hbox{\it scones, tea})$. 
	Its projection onto dimension $1$ is given by $\pi_1(x_{1,2}) = \text{\it scones}$. 
	Similarly, the projection onto dimension $1$ of menu $A_{1,2} = (\{\hbox{\it scones, croissants, cantucci}\}, \{\hbox{\it tea, coffee}\})$ is $\pi_1(A_{1,2}) = A_1 = \{\hbox{\it scones, croissants, cantucci}\}$.
	Finally, consider $c(A_1,A_2) = \{(\hbox{\it scones, tea}), (\hbox{\it cantucci, coffee})\}$. 
	The projection of $c(A_1,A_2)$ contains items in $A_1$ that appear in at least one chosen bundle, that is, $\pi_1(c(A_1, A_2)) = \{\hbox{\it scones, cantucci}\}$.
\end{example}	

Furthermore, we define the family\vspace{-0,1cm}
$$
\pi_S(\mathfrak{M}_Q) = \left\{A_S\in \prod_{q \in S } 2^{X_q} \;\text{ such that }\; A_S=\pi_S(A_Q)\;\text{for some}\; A_Q\in \mathfrak{M}_Q \right\}.
$$

In the next definition, we introduce the notion $S$-separability, a form of independence across dimensions requiring that choices in $S$ solely depend on menus in $S$. 
Separability is then defined by imposing $S$-separability for every nonempty set $S$ of dimensions.

\begin{definition} \label{DEF:separable_joint_choice}
	Let $c \colon \mathfrak{M}_{Q} \to 2^{\prod_{q \in Q} X_q}$ be a joint choice, 
	and fix a nonempty set $S \subseteq Q$ of dimensions.
	Then $c$ is $S$-\textit{separable} if $\pi_S(c(A_S,B_{-S}))=\pi_S(c(A_S,C_{-S}))$ for all $A_S \in \pi_S(\mathfrak{M}_Q)$ and $B_{-S}, C_{-S} \in \pi_{-S}(\mathfrak{M}_Q)$ such that $(A_S,B_{-S}),(A_S,C_{-S})\in\mathfrak{M}_Q$.
	A joint choice on $\mathfrak{M}_{Q}$ is \textit{separable} if it is $S$-separable for all nonempty $S \subseteq Q$. 
\end{definition}

Projections $\pi_S(c(\cdot))$ can be thought of as conditional choices. 
Separability then requires that conditioning on different menus in $-S$ does not affect the choice from dimensions $S$.

\begin{example}[\it Consumption choices, continued]\label{EXAMPLE:consumer_purchase_separability}
	Let $Q = \{1,2,3\}$ index three categories: snacks, beverages and dairy. 
	\begin{itemize}
		\item $X_1 = \{\hbox{\it scones, cantucci}\}$, \quad $X_2 = \{\hbox{\it tea, coffee}\}$, \quad $X_3 = \{\hbox{\it milk, butter}\}$.
	\end{itemize}
	The joint choice $c \colon \mathfrak{M}_Q \to 2^{\prod_{q \in Q} X_q}$ is given by
	\begin{itemize}
		\item $c(A_Q) = c(\{\hbox{\it scones}\}, \{\hbox{\it tea, coffee}\}, \{\hbox{\it milk, butter}\}) = \{(\hbox{\it scones, tea, milk})\}$,
		\item $c(B_Q) = c(\{\hbox{\it cantucci}\}, \{\hbox{\it tea, coffee}\}, \{\hbox{\it milk, butter}\}) = \{(\hbox{\it cantucci, coffee, milk})\}$,
		\item $c(C_Q) = c(\{\hbox{\it scones}\}, \{\hbox{\it tea, coffee}\}, \{\hbox{\it butter}\}) = \{(\hbox{\it scones, tea, butter})\}$.
	\end{itemize}
	To check for $\{3\}$-separability, we use the projection map $\pi_3$ to isolate and compare decisions on dairy across relevant menus:
	\begin{itemize}
		\item $\pi_{3}(c(A_Q)) = \pi_{3}(c(A_{1,2}, \{\hbox{\it milk, butter}\})) =\{\hbox{\it milk}\}$, 
		\item $\pi_{3}(c(B_Q)) = \pi_{3}(c(B_{1,2}, \{\hbox{\it milk, butter}\})) =\{\hbox{\it milk}\}$,
	\end{itemize}
%
	Notice that $\{2\}$-separability does not hold.
	Indeed, we have
		
	\begin{itemize}
		\item $\pi_{2}(c(A_Q)) = \pi_{2}(c(A_{1,3}, \{\hbox{\it tea, coffee}\})) = \{\hbox{\it tea}\}$, 
		\item $\pi_{2}(c(B_Q)) = \pi_{2}(c(B_{1,3}, \{\hbox{\it tea, coffee}\})) = \{\hbox{\it coffee}\}$.
	\end{itemize}

	This is due to complementarities between dimension $1$ and dimension $2$ --- tea is chosen with scones, coffee is chosen with cantucci.
	It follows that $c$ is not separable.

\end{example}	

Definition~\ref{DEF:separable_joint_choice} uses menu variability to formalize $S$-separability.  
In particular, it requires to compare observed joint choices across menus in which only available options for dimensions $-S$ have changed. 
If choices within dimensions $S$ are invariant, $S$-separability is revealed. 
The following result characterizes separability, namely through the existence of `latent' independent choices that explain the whole dataset. 
In particular, $S$-separability requires the existence of a joint choice on the dimensions belonging to $S$, whose picks are independent from the remaining dimensions. 

\begin{lemma} \label{LEMMA:chz_for_separability}
	Let $c \colon \mathfrak{M}_Q \to 2^{\prod_{q \in Q} X_q}$ be a joint choice on $\mathfrak{M}_{Q}\subseteq\prod_{q\in Q} 2^{X_q}$. 
	The following statements are equivalent for any nonempty subset $S$ strictly contained in $Q$: 
	\begin{itemize}
		\item[\rm(i)] $c$ is $S$-separable;
		\item[\rm(ii)] there is a joint choice $c_S \colon \pi_S(\mathfrak{M}_Q) \to 2^{\prod_{q \in S} X_q}$ 
		such that\vspace{-0,1cm}
		$$
		\pi_S \circ c = c_S \circ \pi_S\,,\vspace{-0,1cm} 
		$$ 
		i.e., $\pi_S(c(A_Q)) = c_S(\pi_S(A_Q))$ for all $A_Q \in \mathfrak{M}_Q$.\footnote{Note that the definition of the projection $\pi_S$ differs on the two sides of the equation, because different are the domain and the codomain: see Figure~\ref{FIG:separability}.}
		In this case, we call $c_S$ the \textit{joint choice induced on $S$ by $c$}.
	\end{itemize}
\end{lemma}

\begin{proof}
	\underline{(i)$\implies$(ii):}
	Suppose (i) holds. 
	Define $c_S \colon \pi_S(\mathfrak{M}_Q) \to 2^{\prod_{q \in S} X_q}$ by
	$$
	c_S(A_S)=\pi_S(c(A_Q)) \text{ for some } A_Q\in \mathfrak{M}_Q \text{ with } \pi_S(A_Q)=A_S,
	$$
	for any $A_S \in \pi_S(\mathfrak{M}_Q)$. 
	To prove (ii), it suffices to check that $c_S$ is well-defined.
	That is, for any $A_Q, B_Q \in \mathfrak{M}_Q$ such that $\pi_S(A_Q) = A_S$ and $\pi_S(B_Q) = B_S$, if $A_S = B_S$, then $\pi_S(c(A_Q)) = \pi_S(c(B_Q))$.  
	
	To that end, suppose $A_Q,B_Q \in \mathfrak{M}_Q$ are such that $A_S = \pi_S(A_Q) = \pi_S(B_Q) = B_S$. 
	By definition of $c_S$, $c_S(A_S) = \pi_S(c(A_Q))$ and  $c_S(B_S) =\pi_S(c(B_Q))$. 
	Note that $A_Q=(A_S,A_{-S})$ and $B_Q=(B_S,B_{-S})$ for some $A_{-S}, B_{-S} \in \pi_{-S}(\mathfrak{M}_Q)$.
	Since $A_S=B_S$, condition (ii) yields $\pi_S(c(A_S,A_{-S}))=\pi_S(c(B_S,B_{-S}))$, as claimed.
	\medskip
	 
	\underline{(ii)$\implies$(i):} 
	Fix $A_S \in \pi_S(\mathfrak{M}_Q)$ and $B_{-S},C_{-S} \in \pi_{-S}(\mathfrak{M}_Q)$ such that $(A_S,B_{-S}),(A_S,C_{-S})\in\mathfrak{M}_Q$.
	By hypothesis, there is a joint choice $c_S$ on $\pi_{S}(\mathfrak{M}_Q)$ such that $\pi_S \circ c =  c_S \circ \pi_S$. 
	It follows that
	$$
	\pi_S(c(A_S,B_{-S}))= c_S(\pi_S(A_S,B_{-S}))= c_S(\pi_S(A_S,C_{-S}))=\pi_S(c(A_S,C_{-S})),
	$$
	as claimed. 
\end{proof}

Thus, $S$-separability imposes that isolating dimensions belonging to $S$ in chosen items gives the same outcome as choosing after isolating dimensions in $S$ in the menu.  
The next example serves to illustrate this implication. 

\begin{example}[\it Consumption choices, continued]
	We return to previous example to check for $\{3\}$-separability by verifying that the definition of a choice $c_3$ is possible.
	First, we use the projection map $\pi_3$ to isolate and compare decisions on dairy across relevant menus:
	\begin{itemize}
		\item $\pi_{3}(A_Q) = \{\hbox{\it milk, butter}\}$, \quad  $\pi_{3}(c(A_Q)) = \{\hbox{\it milk}\}$, 
		\item $\pi_{3}(B_Q) = \{\hbox{\it milk, butter}\}$, \quad  $\pi_{3}(c(B_Q)) = \{\hbox{\it milk}\}$,
		\item $\pi_{3}(C_Q) = \{\hbox{\it butter}\}$, \quad $\pi_{3}(c(C_Q)) = \{\hbox{\it butter}\}$.
	\end{itemize}
	This consistency allows us to define $c_3$ so that $c_3(\{\hbox{\it milk, butter}\}) = \{\hbox{\it milk}\}$ and, trivially, $c_3(\{\hbox{\it butter}\}) = \{\hbox{\it butter}\}$. 
	Thus we have 
	\begin{itemize}
		\item $c_3(\pi_3(A_Q)) = \pi_{3}(c(A_Q)) = \{\hbox{\it milk}\}$,
		\item $c_3(\pi_3(B_Q)) = \pi_{3}(c(B_Q)) = \{\hbox{\it milk}\}$,
		\item $c_3(\pi_3(C_Q)) = \pi_{3}(c(C_Q)) = \{\hbox{\it butter}\}$,
	\end{itemize}
	
	which expectedly confirms that the joint choice $c$ is $\{3\}$-separable. 
	A similar procedure verifies that $\{2\}$-separability does not hold.
	Indeed, we have
	
	\begin{itemize}
		\item $\pi_{2}(A_Q) = \{\hbox{\it tea, coffee}\}$, \quad  $\pi_{2}(c(A_Q)) = \{\hbox{\it tea}\}$, 
		\item $\pi_{2}(B_Q) = \{\hbox{\it tea, coffee}\}$, \quad  $\pi_{2}(c(B_Q)) = \{\hbox{\it coffee}\}$,
	\end{itemize}
	
	so there is no $c_2$ such that $c_2(\pi_{2}(B_Q))=\pi_{2}(c(B_Q))=\pi_{2}(c(A_Q))=c_2(\pi_{2}(A_Q))$.
\end{example}

By Definition~\ref{DEF:separable_joint_choice} and Lemma~\ref{LEMMA:chz_for_separability}, a joint choice on $\prod_{q \in Q} X_q$ is always $Q$-separable.
On the other hand, (full) separability is much more demanding, because it assumes in principle the existence of as many as $(2^{\vert Q\vert}\!-\!1)$ independent latent joint choices, which should explain the dataset. 

Figure~\ref{FIG:separability} illustrates the meaning of $S$-separability by means of a commutative diagram in the special case of a complete joint choice, that is, $\mathfrak{M}_{Q} = \prod_{q\in Q} 2^{X_q}$ and $\pi_S(\mathfrak{M}_Q) = \prod_{q\in S} 2^{X_q}$. 


\begin{figure}[h!] 
\begin{center}
\psset{xunit=0.67} \psset{yunit=0.96} 
\begin{pspicture}[showgrid=false](-1,-1.7)(5,5.8)  
\psline[linestyle=solid, linewidth=0.07, arrowsize=7pt]{->}(0.28,4)(4,4)
\psline[linestyle=solid, linewidth=0.07, arrowsize=7pt]{->}(0.28,0)(4,0)
\psline[linestyle=solid, linewidth=0.07, arrowsize=7pt]{->}(-1,3.5)(-1,0.4)
\psline[linestyle=solid, linewidth=0.07, arrowsize=7pt]{->}(5,3.6)(5,0.4)
\rput(2,2){\huge $\circlearrowleft$}
\rput(-1.1,4){\large $\prod_{q\in Q} 2^{X_q}$} 
\rput(5.1,4.07){\large $2^{^{\prod_{q\in Q} X_q}}$} 
\rput(-1.1,0){\large $\prod_{q\in S} 2^{X_q}$} 
\rput(5.1,0.07){\large $2^{^{\prod_{q\in S} X_q}}$}
\rput(-3,4.8){$A_Q$} 
\rput(7.2,4.8){$c(A_Q)$}
\rput(-2.7,-1){$\pi_S(A_Q)=A_S$}
\rput(5.7,-1){$c_S(A_S)=\pi_S(c(A_Q))$}
\rput(2,4.37){\Large$c$} \rput(2,-0.4){\Large{$c_S$}}
\rput(-1.75,2){\Large $\pi_S$}  \rput(5.8,2){\Large{$\pi_S$}}
\pscurve[linestyle=solid,linewidth=0.025,arrowsize=6pt]{->}(-2.2,4.9)(2,5.3)(6.1,4.9)
\pscurve[linestyle=solid,linewidth=0.025,arrowsize=6pt]{->}(-2.2,-1.35)(2,-1.8)(6.2,-1.35)
\pscurve[linestyle=solid,linewidth=0.025,arrowsize=6pt]{->}(-3,4.4)(-3.7,2)(-3,-0.65)
\pscurve[linestyle=solid,linewidth=0.025,arrowsize=6pt]{->}(6.8,4.4)(7.5,2)(6.8,-0.65)
\end{pspicture}
\end{center}
\caption{The $S$-separability of a joint choice $c$ in terms of a commutative diagram.\label{FIG:separability}}
\end{figure}


	$S$-separability naturally applies to consumer purchasing behavior, particularly because empirical data often manifest as joint choices.
	This is the case, for instance, of consumption scanner data adopted in \cite{deHaanvanderGrient2011} and  \cite{RenkinMontialouxSiegenthaler2022}.
	Further instances of $S$-separability are provided in different settings.
	
	Let indices in $Q$ represent time periods, so that the choice at each $q$ picks consumption at period $q$ and the joint choice selects streams of consumption. 
	An agent may consider the first $S$ periods as `short term', and group the remaining dimensions $-S$ as `long term'. 
	The choice is then made by maximizing a utility function of the form $u(c_1, \dots, c_Q) = U(V_S(c_1, \dots, c_S) + V_{-S}(c_{S + 1}, \dots, c_Q))$, where $V_S$ represents preferences during working time and $V_{-S}$ preferences during retirement. 
		The resulting choice is $S$-separable.\footnote{More generally, we will show that any additively separable utility function  induces a separable choice.}
		Streams of consumptions may also resemble \textit{dated rewards}, i.e., temporal sequences of payments, which are the alternatives faced by experimental subjects in the so-called \textit{multiple-price list method} \citep{Binswanger1980,AndersenHarrisonLauRutstrom2006,AndersenHarrisonLauRutstrom2007}.     
		Dated rewards are also the primitives of the intertemporal choice models of \cite{CollerWilliams1999}, \cite{ToubiaJohnsonEvgeniouDelquie2012}, \cite{Blavatskyy2017}, and \cite{Pennesi2021}, among others.  

	Alternatively, let $Q$ be a set of choices over lotteries, and consider a `narrow bracketer' \citep{RabinWeizsacker(2009)}, who performs the choice at each dimension separately, picking the lottery that maximizes a given preference. 
	With such procedure, the agent ignores all correlations between lotteries that appear in different choices, so that separated decisions may generate suboptimal global outcomes.
	As the agent fails to consider how options at each dimension combine with the rest, separability arises.\footnote{For further experimental evidence of choice bracketing, we refer to \cite{EllisFreeman2024}.}
Finally, indices in $Q$ can represent different tasks, each performed by a department of a firm, or a member of a household. 
	The way in which each decision maker allocates its budget for each task may or may not depend on choices made by other decision makers.\footnote{This application is actually suggested in \cite{Strotz1957} for separable utility functions.}
	For the organization of the firm, $S$ and $T$-separability, for $S,T \subseteq Q$, yields that funding decisions for tasks in $S$ and $T$ can be performed by separate departments with no inefficiencies. 

\section{Stability and characterization of separability} \label{SECT:separability_behaves_well}

In this section we examine several features of separability.
Specifically, first 
we show that separability of a joint choice is stable under the operations of taking unions and intersections of dimensions.\footnote{For the latter, we shall need a condition of richness of the choice domain.}
While insightful per se, these results are ancillary to the rest of our analysis. 
Next, 
we provide a simple characterization, and show that the separability can distinguish a class of choice models and allocation of resources.


\subsection*{Stability under union and intersection} \label{SUBSECT:stability}

We start with a negative (but expected) result: a $S$-separable joint choice needs  not be separable with respect to subsets or supersets of $S$. 
We show this in the next counterexample. 

\begin{counterexample}[\it Non-stability of separability for subsets and supersets]\label{EX:separability_not_stable_inclusion}
	 On $Q=\{1,2,3\}$, let $X_1=ab$, $X_2=xy$, and $X_3=pq$.
	 Define a joint choice on $\mathfrak{M}_Q= \{(ab,xy,p),(ab,xy,q)\}$ as follows:
	 $$
	 c \colon \mathfrak{M}_{1,2,3}\to 2^{X_1\times X_2 \times X_3} \,,\quad
	 c(ab,xy,p)=\{(a,x,p),(b,x,p)\} \quad \text{and} \quad c(ab,xy,q)=\{(a,x,q)\}\,.
	 $$
	Definition~\ref{DEF:separable_joint_choice} implies that $c$ is $Q$-separable.
	However, $c$ fails to be $\{1\}$-separable, because a choice induced on $\{1\}$ by $c$ does not exist.
	Furthermore, $c$ is $\{2\}$-separable, since $\pi_2(c(ab,xy,p)) = \pi_2(c(ab,xy,q) = x$. 
	Finally, $c$ is $\{1,2\}$-separable: indeed, by Lemma~\ref{LEMMA:chz_for_separability}, in order not to be $S$ separable,  there must be at least two distinct sets having the same $S$-projection, and  this is false for $S=\{1,2\}$.
%
%
\end{counterexample}

On the contrary, $S$-separability behaves well with respect to elementary set-theoretic binary operations, provided that the choice domain satisfies a closure property.

\begin{definition}\label{DEF:menus_betweenness}
	Let $S,T\subseteq Q$ two nonempty sets of dimensions. 
	A family $\mathfrak{M}_{Q}\subseteq\prod_{q\in Q} 2^{X_q}$ satisfies \textit{menus betweenness with respect to $S$ and $T$} if for any $A_Q,B_Q\in\mathfrak{M}_Q$ such that $\pi_{S\cap T}(A_Q)=\pi_{S\cap T}(B_Q)$, there is $E_Q\in\mathfrak{M}_Q$ such that $\pi_{S}(E_Q)=\pi_{S}(A_Q)$ and $\pi_{T}(E_Q)=\pi_{T}(B_Q)$.
\end{definition}

In other words, given two sets of dimensions $S$ and $T$, menus betweenness with respect to them means that for every feasible pair of menus $A_{Q},B_{Q}$ agreeing on the dimensions in $S\cap T$, there is some available menu $E_Q$ which agrees with $A_Q$ on $S$ and with $B_Q$ on $T$.
This richness property ensures stability of $S$-separability under intersections of subsets of dimensions:

\begin{theorem}\label{THM:separability_with_union_and_intersection}
	Let $c \colon \mathfrak{M}_{Q} \to 2^{\prod_{q \in Q} X_q}$ be a joint choice on $\mathfrak{M}_{Q}\subseteq\prod_{q\in Q}2^{X_q}$, and $S,T \subseteq Q$ nonempty subsets of dimensions. 
	If $c$ is both $S$-separable and $T$-separable, then the following statements hold: 
	\begin{itemize}
		\item[\rm (i)] $c$ is $(S \cup T)$-separable;
		\item[\rm (ii)] $c$ is $(S\cap T)$-separable, provided that $\mathfrak{M}_Q$ satisfies menus betweenness with respect to $S$ and $T$.
	\end{itemize}
\end{theorem} 	 	

\begin{proof}
	Suppose $c$ is $S$-separable and $T$-separable.
	\smallskip
	
	\underline{(i)} 
	 Fixed $A_Q \in \mathfrak{M}_{Q}$, we have to show that there is $c_{S \cup T}\colon \pi_{S\cup T}(\mathfrak{M}_{Q})\to 2^{\X_{S \cup T}}$ such that the equality $\pi_{S \cup T}(c(A_Q))=c_{S \cup T}(\pi_{S \cup T}(A_Q))$ holds.
	Define $c_{S \cup T}(\pi_{S\cup T}(A_{Q}))=\pi_{S \cup T}(c(D_Q))$, where $D_Q \in  \mathfrak{M}_{Q}$ is such that $\pi_{S \cup T}(D_Q)=A_{S \cup T}$.
	Note that such $D_Q$ always exists, since we can consider $D_Q=A_Q.$
	To conclude, we show that $c_{S \cup T}$ is well-defined.
	Let $A_Q, B_Q \in  \mathfrak{M}_{Q}$ be such that $\pi_{S \cup T}(A_Q)=\pi_{S \cup T}(B_{Q})$, we show that $\pi_{S \cup T}(c(A_Q))=\pi_{S \cup T}(c(B_Q))$.
	Then $S$-separability and $\pi_{S \cup T}(A_Q)=\pi_{S \cup T}(B_{Q})$  imply 
	$$
	\pi_S(c(A_Q))=c_S(\pi_S(A_Q))=c_S(\pi_S(B_Q))=\pi_S(c(B_Q)),
	$$
	whereas $T$-separability and $\pi_{S \cup T}(A_Q)=\pi_{S \cup T}(B_{Q})$ yield
	$$
	\pi_T(c(A_Q))=c_T(\pi_T(A_Q))=c_T(\pi_T(B_Q))=\pi_T(c(B_Q)).
	$$
	We conclude that $\pi_{S \cup T}(c(A_Q))=\pi_{S \cup T}(c(B_Q))$.
\medskip

\underline{(ii)}
Assume now that $\mathfrak{M}_Q$ satisfies menus betweenness with respect to $S$ and $T$.
 Fixed $A_Q \in\mathfrak{M}_Q$, we show that there is a joint choice $c_{S \cap T} \colon \pi_{S\cap T}(\mathfrak{M}_{Q}) \to 2^{\X_{S \cap T}}$ such that $\pi_{S \cap T}(c(A_Q))=c_{S \cap T}(\pi_{S \cap T}(A_Q))$.
	Define $c_{S \cap T}(\pi_{S\cap T}(A_Q))=\pi_{S \cap T}(c(D_Q))$, where $D_Q \in \mathfrak{M}_{Q}$ is such that $\pi_{S \cap T}(D_Q)=A_{S \cap T}$.
	Again, such $D_Q$ always exists, since we can take $D_Q=A_Q.$
	To conclude the proof, it suffices to show that $c_{S \cap T}$ is well-defined.
	Suppose $A_Q$ and $B_Q$ are menus in  $\mathfrak{M}_{Q}$ such that $\pi_{S \cap T}(A_Q)=\pi_{S \cap T}(B_Q)$.
	 We need to prove that $c_{S \cap T}(\pi_{S \cap T}(A_Q))=c_{S \cap T}(\pi_{S \cap T}(B_Q))$.
	Let $E_Q \subseteq X_Q$ be such that $\pi_S(E_Q)=\pi_S(A_Q)$ and $\pi_{T}(E_Q)=\pi_{T}(B_Q)$.
	Since $\mathfrak{M}_Q$ satisfies menus betweenness with respect to $S$ and $T$, we know that $E_Q\in\mathfrak{M}_Q$.  
	By $S$-separability and the equality $\pi_S(A_Q)=\pi_S(E_Q)$, we get
	$$
	\pi_S(c(A_Q))=c_S(\pi_S(A_Q))=c_S(\pi_S(E_Q))=\pi_S(c(E_Q)).
	$$
	Furthermore, $T$-separability and the equality $\pi_{T}(B_Q)=\pi_{T}(E_Q)$ imply that
	$$
	\pi_T(c(B_Q))=c_T(\pi_T(B_Q))=c_T(\pi_T(E_Q))=\pi_T(c(E_Q)).
	$$
	Since $\pi_{S \cap T}(\pi_S(c(E_Q)))=\pi_{S \cap T}(\pi_T(c(E_Q)))$, we finally conclude 
	$$
	\pi_{S \cap T}(c(A_Q))=\pi_{S \cap T}(\pi_S(c(A_Q)))=\pi_{S \cap T}(\pi_T(c(B_Q)))=\pi_{S \cap T}(c(B_Q)),
	$$
	as claimed.
\end{proof}

An immediate consequence of Theorem~\ref{THM:separability_with_union_and_intersection} is the following:

\begin{corollary} \label{COR:stability_for_complete_choices}
	A complete joint choice $c$ that is $S$-separable and $T$-separable is also $(S\cup T)$-separable and $(S\cap T)$-separable.
	Therefore, the family of all subsets $S$ of $Q$ for which $c$ is $S$-separable forms a bounded lattice under set-inclusion, with the operations of intersection and union.
\end{corollary}

The next counterexample shows that if menus betweenness with respect to $S$ and $T$ does not hold, then $(S\cap T)$-separability may fail. 

\begin{counterexample}[\it Non-stability of separability for intersection] 
	For $Q=\{1,2,3\}$, let 
	\begin{itemize}
		\item $X_1=ab$, $X_2=pq$, $X_3=xy$,
		\item $\mathfrak{M}_Q= \{(ab,pq,x),(a,pq,xy)\} =\{A_Q,B_Q\}$,
		\item $S=\{1,2\}$, $T=\{2,3\}$.
	\end{itemize}
	Note that $\mathfrak{M}_Q$ does not satisfy menus betweenness with respect to $S$ and $T$, because $\pi_{S \cap T} (A_Q) = \pi_{S \cap T}(B_Q)$ holds, and yet there is no $E_Q \in \mathfrak{M}_Q$ such that $\pi_S(E_Q)=\pi_S(A_Q)$ and $\pi_T(E_Q)=\pi_T(B_Q)$. 
	 Now define a joint choice $c$ on $\mathfrak{M}_Q$ as follows:
	 $$
	 c\colon \mathfrak{M}_Q \to 2^{X_{1}\times X_{2}\times X_{3}}\,,\quad c(ab,pq,x)=\{(b,p,x)\} \quad \text{and} \quad c(a,pq,xy)=\{(a,q,y)\}\,.
	 $$ 
	It is easy to check that $c$ is both $S$-separable and $T$-separable, but fails to be $(S \cap T)$-separable.
%
\end{counterexample}

The following natural extension of Theorem~\ref{THM:separability_with_union_and_intersection} will be useful to justify the approach undertaken in Section~\ref{SECT:testing_separability}.

\begin{corollary}\label{COR:separability_with_finite_union_and_finite_intersection}
	Let $c \colon \mathfrak{M}_{Q} \to 2^{\prod_{q \in Q} X_q}$ be a joint choice on $\mathfrak{M}_{Q}\subseteq\prod_{q\in Q}2^{X_q}$, and $S_1, S_2, \ldots, S_n$ nonempty subsets of $Q$. 
	If $c$ is $S_j$-separable for all $j \in \{1,2,\ldots,n\}$, then we have:
	\begin{itemize}
		\item[\rm (i)] $c$ is $(S_1 \cup S_2 \cup \ldots \cup S_n)$-separable;
		\item[\rm (ii)] $c$ is $(S_1 \cap S_2 \cap \ldots \cap S_n)$-separable, provided that there exists a permutation $\sigma$ of $\{1,2,\ldots,n\}$ such that $\mathfrak{M}_Q$ satisfies menus betweenness with respect to the following $n\!-\!1$ pairs of sets:\vspace{-0,4cm}
		\begin{align*}
			S_{\sigma(1)} & \quad \text{and} \quad S_{\sigma(2)},\\
			S_{\sigma(1)} \cap S_{\sigma(2)} & \quad \text{and} \quad S_{\sigma(3)},\\
			&\ldots\\ 
			S_{\sigma(1)} \cap S_{\sigma(2)} \cap \ldots \cap S_{\sigma(n-1)}& \quad \text{and} \quad S_{\sigma(n)}.
		\end{align*}  
	\end{itemize}
\end{corollary} 	 	

\begin{proof}
	Part (i) is proved by induction, using Theorem~\ref{THM:separability_with_union_and_intersection}(i). 
	Part (ii) follows from Theorem~\ref{THM:separability_with_union_and_intersection}(ii) by induction, upon observing that the operation of intersection is associative.\footnote{For instance, we read $S_{\sigma(1)} \cap S_{\sigma(2)} \cap S_{\sigma(3)}$ as $\left(S_{\sigma(1)} \cap S_{\sigma(2)}\right) \cap S_{\sigma(3)}$, etc.} 
\end{proof}

\subsection*{Characterization and applications}\label{SUBSECT:characterization_separability} 

Theorem~\ref{THM:separability_with_union_and_intersection} yields a useful characterization of separability.

\begin{theorem}\label{THM:q_separability_equivalent_separability_by_dimensions}
	The following statements are equivalent for a joint choice function $c$ on $\mathfrak{M}_{Q}\subseteq\prod_{q\in Q} 2^{X_q}$:
	\begin{itemize}
		\item[\rm(i)] $c$ is separable; 
		\item[\rm(ii)] $c$ is $\{q\}$-separable for all $q \in Q$;
		\item[\rm(iii)] there is a family $\left(c_q \colon\pi_{q}(\mathfrak{M}_Q)\to X_q \right)_{q \in Q}$ of choice functions respectively defined on $\pi_{q}(\mathfrak{M}_Q)\subseteq 2^{X_q}$ such that $c(A_Q)=(c_q(\pi_{q}(A_Q)))_{q\in Q}=(c_q(A_q))_{q \in Q}$ for any $A_Q\in \mathfrak{M}_Q$.
	\end{itemize}
\end{theorem}

\begin{proof}
	Statements (i) and (ii) are equivalent by virtue of Theorem~\ref{THM:separability_with_union_and_intersection}(i).
	Moreover, statement (iii) obviously implies statement (ii).
	Therefore, it suffices to show that (ii) implies (iii). 
	Assume that $c \colon \mathfrak{M}_Q \to \prod_{q \in Q} X_q$ is $\{q\}$-separable for all $q \in Q$.
	To prove the claim, we have to show that for all $q \in Q$, there is a choice function $c_q$ on $\pi_q(\mathfrak{M}_Q)$ such that $c(A_{Q}=\prod_{q \in Q}A_q)=(c_q(A_q))_{q \in Q}$ for all $A_Q=\prod_{q \in Q} A_q \in \mathfrak{M}_Q$.
	Fix $q \in Q$.
	By $\{q\}$-separability, there is a choice function $c_q$ on $\pi_q(\mathfrak{M}_Q)$ such that $\pi_q(c(A_Q))=c_q(\pi_q(A_Q))=c_q(A_q)$ for any $A_Q\in \mathfrak{M}_Q$.
	Note that $c(A_Q)=(\pi_q(c(A_Q)))_{q \in Q}=(c_q(A_q))_{q \in Q}$ for any $A_Q\in \mathfrak{M}_Q$.
\end{proof}

The equivalence between (i) and (ii) in Theorem~\ref{THM:q_separability_equivalent_separability_by_dimensions} yields an important economic consequence: if one-dimensional picks are independent from the remaining dimensions, then the joint choice induced by any aggregation of dimensions is not affected by options available in the other dimensions.

Theorem~\ref{THM:q_separability_equivalent_separability_by_dimensions} also illustrates how we naturally generalize a distinct formulation of separability  introduced by \cite{ChambersMasatliogluTuransick2024}, which applies to single valued joint choices over two-dimensional alternatives --- thus generated by two independent choice functions.\footnote{Formally, in \cite{ChambersMasatliogluTuransick2024} a \textit{$2$-joint choice function on} $2^{X_1 \times X_2}$ is a map $c \colon 2^{X_1} \times 2^{X_2} \to X_1 \times X_2$ such that $c(A,B) \in A \times B$ for all $A \in 2^{X_1}$ and $B \in 2^{X_2}$. 
	Then $c$ is \textit{separable} if there are two choice functions $c_1 \colon 2^{X_1} \to X_1$ and $c_2 \colon 2^{X_2} \to X_2$ such that $c(A,B) = (c_1(A), c_2(B))$ for all $A \in 2^{X_1}$ and $B \in 2^{X_2}$.} 
Rather, separable joint choice functions in our definition are multidimensional datasets, and the equivalence between parts (i) and (iii) in Theorem~\ref{THM:q_separability_equivalent_separability_by_dimensions} shows that separable joint functions are generated by $\vert Q\vert$-many distinct choice functions.

There is a further connection between separability of joint choices and choices on mono-dimensional alternatives.
Indeed, behavioral assumptions determining models of deterministic choice can inherently generate separable or non-separable choices.

\begin{example}[\it Rational choices]\label{EXAMPLE:rational_choice}
	Let $Q$ index decisions in a sequence, fix a finite set of items $X$, and let $X_q = X$ for each $q$. 
	Menus in $\mathfrak{M}_Q$ are vectors of subsets of $X$. 
	A rational agent is endowed with a strict preference $\succsim$ on $X$.\footnote{For the moment, although they are well known to readers, we do not formally introduce individual preferences.
	A rigorous definition will be provided in Section \ref{SECT:rational_vs_separable}.} 
	At each decision $q$, the choice is made by selecting maximal items in $A_q$ with respect to $\succsim$, i.e. $\max(A_q, \succsim)$. 
	Thus, $c(A_Q) = c(A_1, \dots, A_{\vert Q\vert}) = (\max(A_1, \succsim~), \max(A_2, \succsim~),\dots, \max(A_{\vert Q\vert}, \succsim))$ holds. 
	Clearly, no interdependencies across dimensions can arise, and joint choice $c$ is separable. 	
\end{example}

A rational agent chooses by maximizing her preference, regardless of past and future choices. 
This yields separability.
Yet, separability is not tied to rationality. 

\begin{example}[\it Choices with limited attention]\label{EXAMPLE:limited_attention}
	As above, let $Q$ index decisions in a sequence, fix a finite set of items $X$ and let $X_q = X$ for each $q$, so that menus in $\mathfrak{M}_Q$ are vectors of subsets of $X$. 
	The agent is endowed with a strict preference $\succ$ on $X$, and, as proposed by \cite{MasatliogluNakajimaOzbay2012}, an attention filter $\Gamma: 2^X \to 2^X$, associating to each set a subset of noticed items. 
	At each decision $q$, the choice is made by taking $\max(\Gamma(A_q),\succ)$, that is, by maximizing the preference within the consideration set. 
	It follows that $c(A_Q) = c(A_1, \dots, A_{\vert Q\vert}) = (\max(\Gamma(A_1), \succ), \max(\Gamma(A_2), \succ), \dots, \max(\Gamma(A_{\vert Q\vert}), \succ))$. 
	Again, $c$ is separable, though choices $c_q$ are not rationalizable.	
\end{example}

Notice that classical notions of separability are not applicable to Example~\ref{EXAMPLE:limited_attention}. 
This is exactly due to the lack of a preference or utility representation for choices $c_q$. 
See Section~\ref{SECT:rational_vs_separable} for further comments.\\

Violations of separability can arise whenever past choices serve as a status quo \citep{MasatliogluOk2005} or a reference point \citep{ApesteguiaBallester2009}, when memory plays a role or attention is dynamic \citep{Rozen(2010), Cerigioni2021}.

\begin{example}[\it Choices with status quo bias]\label{EXAMPLE:choice_sequences_statusquo}
	We observe the behavior of an individual in two subsequent choice problems, $Q = \{1, 2\}$, and let $X_1 = X_2 = xyz$. 
	Items $x$ and $y$ are strictly preferred to $z$, while there is no clear ranking between $x$ and $y$. 
	Besides preferences, choice behavior is informed by a \textit{status quo bias}, as in \cite{MasatliogluOk2005}, entailing that the agent experiences some psychological discomfort when deviating from the last chosen option (status quo). 
	Specifically, whenever the status quo alternative is available and not clearly dominated, it is selected. 
	In the first sequence, let $A_1 =xz$ and $A_2 =xy$. 
	The selection is $c(xz,xy) = \{(x, x)\}$, because the agent maximizes the preference in the first choice problem, and then sticks to the status quo. 
	In the second sequence, let $B_1 =yz$ and $B_2 = A_2 =xy$. 
	The selection is now $c(yz,xy) = \{(y,y)\}$, where the agent follows the same criterion. 
	Note how the pick from the (mono-dimensional) menu $xy$ is influenced by previous choices. 
	Any joint choice $c$ such that  $c(xz,xy) = (x, x),$ and $c(yz, xy) = (y, y)$ is not $\{2\}$-separable, because $\pi_{2}(c(A_1,A_2))=x \neq y =\pi_{2}(c(B_1,A_2))$.
\end{example}

When $Q$ represents individuals in an economy, separability is a property of allocation of resources. 

\begin{example}[\it Efficient and fair allocations of resources]\label{EXAMPLE:efficient_allocation_resources}
	Consider an economy with two individuals, $Q = \{1, 2\}$ and $X_1 = X_2 = ab$, both strictly preferring item $a$ over $b$ (that is, $a \succ_1 b$ and $a \succ_2 b$).
	The complete joint choice $c^E$ that always assigns the most preferred item results in a separable joint choice. 
	Indeed, we have:
	\begin{itemize}
	\item $\pi_{1}(c(a,a))=\pi_{1}(c(a,b))=\pi_{1}(c(a,ab))=\{a\},$
	$\pi_{1}(c(b,a))=\pi_{1}(c(b,b))=\pi_{1}(c(b,ab))=\{b\},$
 $\pi_{1}(c(ab,a))=\pi_{1}(c(ab,b))=\pi_{1}(c(ab,ab))=\{a\},$ and
 \item $\pi_2(c(a,a))=\pi_{2}(c(b,a))=\pi_{2}(c(ab,a))=\{a\},$ $\pi_2(c(a,b))=\pi_{2}(c(b,b))=\pi_{2}(c(ab,b))=\{b\},$  $\pi_2(c(a,ab))=\pi_{2}(c(b,ab))=\pi_{2}(c(ab,ab))=\{a\}.$
	\end{itemize}
%

A failure of separability in the distribution rule can instead reveal equity concerns.


	Consider an economy with two individuals, $Q = \{1, 2\}$ and $X_1 = X_2 = ab$, both strictly preferring item $a$ over $b$ (that is, $a \succ_1 b$ and $a \succ_2 b$).
	Suppose that resources are assigned according to a fairness criterion, in particular by selecting envy-free allocations whenever possible.\footnote{Recall that an allocation $(a_1, a_2)$ is \textit{envy-free} if $a_1 \succsim_1 a_2$ and $a_2 \succsim_2 a_1$. More generally, an allocation $(a_1, \dots, a_{\vert Q\vert})$ is \textit{envy-free} if any agent $i \in Q$ weakly prefers the allocated good $a_i$ to $a_j$, for all $j \in Q$. See \cite{Varian1974}.} 
	Thus, $c^F(ab,ab) = \{(a, a)\}$ and $c^F(ab,b) = \{(b, b)\}$. 
	Then, since $\pi_{1}(c^F(ab,ab)) = a \neq b = \pi_{1}(c^F(ab,b))$, it follows that $c^F$ is not separable.	
\end{example}

Not all forms of fairness entail non-separability, consider for instance \textit{sufficientarianism} \citep{AlcantudMariottiVeneziani2022}. 
Finally, in multi-agent choices where the outcome specifies a chosen option per individual in a group, a failure of separability indicates a degree of choice complementarity or social influence  \citep{Cuhadaroglu2017,DemirkanBoyao2022}. 


Thus, choice separability sheds light on two distinct aspects. 
First, it can uncover complementarities inherent in the objects of choice.
Second, it reveals key features of the cognitive processes and criteria involved in decision-making.
The next section provides useful insights for the design of parsimonious separability tests.


\section{Selective families and separability tests}\label{SECT:testing_separability}

In this section we use some implications of Theorem~\ref{THM:separability_with_union_and_intersection} to simplify the verification of separability.
In particular, we prove that whenever the choice domain satisfies a suitable property, the amount of computations needed to assess separability can be significantly lessened.


By Theorem~\ref{THM:q_separability_equivalent_separability_by_dimensions}, separability of a joint choice function can be elicited from $\{q\}$-separability of the dataset for any dimension $q\in Q$.
However, as soon as the cardinality of $Q$ becomes high, the assessment of $\{q\}$-separability for any $q\in Q$ becomes rather problematic.
To effectively address this issue, next we introduce special families of subsets of dimensions. 

%
%
\begin{definition} \label{DEF: Martin family}
	 A family $\mathscr{S}=\{S_i: i \in I\}$ of distinct nonempty subsets of $Q$ is \textit{selective (for $Q$)} if for every $q \in Q$, there is a nonempty set $I(q) \subseteq I$ such that $\{q\}=\bigcap \{S_i : i \in I(q)\}$. 
Equivalently, $\mathscr S$ is selective if for any $q\in Q$ there is a family $\mathscr S_q$ of nonempty subsets of $\mathscr S$ such that $\{q\} =\bigcap \mathscr S_q$.
		The smallest size of a selective family for $Q$ is denoted by $\mathrm{sel}(Q)$.  
\end{definition}

Note that the definition of $\mathrm{sel}(Q)$ is sound, because there is always a selective family for $Q$, namely the collection of singletons; thus, $\mathrm{sel}(Q)\leqslant \vert Q\vert$.  
For instance, when $Q =\{1,2,3,4,5,6\}$, then $\mathrm{sel}(Q) =4$.
 Indeed, the family $\mathcal S = \big\{\{1,2,3\},\{1,4,5\},\{2,4,6\},\{3,5,6\}\big\}$ is selective for $Q$, and there is no selective family of smaller size. 

The usefulness of a selective family in detecting the separability of a joint choice is connected to the synergic action of   Theorem~\ref{THM:q_separability_equivalent_separability_by_dimensions} and Corollary~\ref{COR:separability_with_finite_union_and_finite_intersection}(ii). 
In fact, Theorem~\ref{THM:q_separability_equivalent_separability_by_dimensions} ensures that separability is equivalent to $q$-separability for all $q \in Q$, whereas Corollary~\ref{COR:separability_with_finite_union_and_finite_intersection}(ii) implies that each $q$-separability can be deduced --- under suitable conditions --- from the $S_j$-separability of all elements in $\mathscr S_q$. 
One last element is missing.
In fact, to allow the experimenter to deduce separability only by verifying that the joint choice is $S_i$-separable for any $S_i$ in a selective family $\mathscr {S}$, we need a property that guarantees the possibility to use Corollary~\ref{COR:separability_with_finite_union_and_finite_intersection}(ii).

\begin{definition}\label{DEF:chained_menus_betweennes}
Let $Q$ be a set of dimensions, and $\mathscr S= \{S_i: i \in I\}$ a selective family for $Q$.
For any $q\in Q$, let $\mathscr{S}_q$ be a family of subsets of $\mathscr S$ such that $\{q\}= \bigcap \mathscr S_q$.
$\mathfrak{M}_{Q}\subseteq  \prod_{q \in Q} 2^{X_q}$ satisfies \textit{$\mathscr S$-betweenness} if for any $q\in Q$, there is a labelling $\{S_{q,1},S_{q,2},\ldots, S_{q,\vert I(q) \vert}\}$ of the elements of $\mathscr S_q$ such that $\mathfrak{M}_{Q}$ satisfies menus betweenness with respect to the following $(\vert I(q) \vert -1)$ pairs of sets:\vspace{-0,4cm}
		\begin{align*}
			S_{q,1} & \quad \text{and} \quad S_{q,2}\,,\\
			S_{q,1} \cap S_{q,2} & \quad \text{and} \quad S_{q,3}\,,\\
			&\ldots\\ 
			S_{q,1} \cap S_{q,2} \cap \ldots \cap S_{q,\vert I(q) \vert -1}& \quad \text{and} \quad S_{q,\vert I(q) \vert}\,.
		\end{align*}  
\end{definition}

	To illustrate, fix $Q=\{1,2,3\}$.
	The family $\mathscr{S}\!=\!\{\{1,2\},\{2,3\},\{1,3\}\}$ is selective for $Q$, with $\mathscr S_1\!=\! \{\{1,2\},\{1,3\}\}$, $\mathscr S_2\!=\! \{\{1,2\},\{2,3\}\}$, and $\mathscr S_3\!=\! \{\{1,3\},\{2,3\}\}$.
	Let $X_1=a$, $X_2=bcd$, $X_3=ef$.
	We prove that the set of $3$-dimensional menus
	$$
	\mathfrak{M}_Q=\{(a,c,e), (a,b,f), (a,c,f)\}
	$$
	satisfies $\mathscr S$-betweenness.	
	To that end, we show that $\mathfrak{M}_Q$ satisfies menu betweenness with respect to the elements of $\mathscr S_1$, those of $\mathscr S_2$, and those of $\mathscr S_3$.
	For $\mathscr S_1$, we have three cases to be considered.
   	\begin{itemize}
		\item If $\pi_1(a,c,e)=\pi_{1}(a,b,f)$, then we need to find $E_Q \in \mathfrak{M}_Q$ such that $\pi_{1,2}(E_Q)=\pi_{1,2}(a,c,e)$ and $\pi_{1,3}(E_Q)=\pi_{1,3}(a,b,f)$. Since $\pi_{1,2}(a,c,f)=\pi_{1,2}(a,c,e)$ and $\pi_{1,3}(a,c,f)= \pi_{1,3}(a,b,f)$, $E_Q=(a,c,f)$ satisfies the claim. 
		\item If $\pi_1(a,c,e)=\pi_1(a,c,f)$, then we can set $E_Q=(a,c,f)$, because $\pi_{1,2}(a,c,f)=\pi_{1,2}(a,c,e)$ and $\pi_{1,3}(a,c,f)=\pi_{1,3}(a,c,f)$.
		\item Finally, for $\pi_1(a,b,f)=\pi_1(a,c,f)$, set $E_Q=(a,b,f)$. 
	\end{itemize} 
	For $\mathscr S_2$, observe that the only case to be considered is $\pi_2(a,c,e)=\pi_2(a,c,f)$. Since $\pi_{1,2}(a,c,f)=\pi_{1,2}(a,c,e)$ and $\pi_{2,3}(a,c,f)=\pi_{2,3}(a,c,f)$, we can take $E_Q = (a,c,f)$. 
	Finally, for $\mathscr S_3$, note that $\pi_3(a,b,f)=\pi_3(a,c,f)$, $\pi_{1,3}(a,c,f)=\pi_{1,3}(a,b,f)$, and $\pi_{2,3}(a,c,f)=\pi_{2,3}(a,c,f)$, so we can take $E_Q = (a,c,f)$. \\

As announced, the following sufficient condition for separability holds:
 
\begin{theorem} \label{THM:separability_by_selective_families}
	Let $c \colon \mathfrak{M}_{Q} \to 2^{\prod_{q \in Q} X_q}$ be a joint choice on $\mathfrak{M}_{Q}\subseteq\prod_{q\in Q}2^{X_q}$, and $\mathscr S= \{S_i: i \in I\}$ a selective family for $Q$.
	If $\mathfrak{M}_{Q}$ satisfies $\mathscr S$-betweenness and
	 $c$ is $S_i$-separable for all $i \in I$, then $c$ is separable.
\end{theorem}

\begin{proof}
For any $q\in Q$, let $\mathscr{S}_q$ be a family of subsets of $\mathscr S$ such that $\{q\}= \bigcap \mathscr S_q$.
	Since $\mathfrak{M}_Q$ satisfies $\mathscr S$-betweenness, for any $q\in Q$,
	there is a labelling $\{S_{q,1},S_{q,2},\ldots, S_{q,\vert I(q) \vert}\}$ of the elements of $\mathscr S_q$ such that $\mathfrak{M}_{Q}$ satisfies menus betweenness with respect to the $(\vert I(q) \vert -1)$ pairs of sets given in Definition~\ref{DEF:chained_menus_betweennes}. 
	Now Corollary~\ref{COR:separability_with_finite_union_and_finite_intersection}(ii) gives $q$-separability for each $q \in Q$, and Theorem~\ref{THM:q_separability_equivalent_separability_by_dimensions} yields full separability of $c$. 
\end{proof}

Since the completeness of the domain of a joint choice guarantees $\mathscr S$-betweenness for any selective family $\mathscr S$, Theorem~\ref{THM:separability_by_selective_families} readily yields 

\begin{corollary}
	For any complete joint choice $c$ on $\mathfrak{M}_{Q}$, if $\mathscr S= \{S_i: i \in I\}$ is a selective family for $Q$, and $c$ is $S_i$-separable for all $i \in I$, then $c$ is separable.   
\end{corollary}

We conclude this section with a combinatorial result, which may allow to significantly reduce the amount of computations needed to assess separability. 

\begin{lemma}\label{LEMMA: smallest Martin family}
    If $n$ is the least integer such that $\vert Q \vert \leq$ $n \choose \lfloor \frac{n}{2} \rfloor$, then $\mathrm{sel}(Q) = n$.
\end{lemma}

\begin{proof}
    Let $\{S_1, \ldots , S_n\}$ be a selective family for $Q$. 
	By definition, the family  $\{I(q): q \in Q\}$ is a Sperner family.\footnote{A \textit{Sperner family} is a collection of subsets of a given set such that none of them is included in another one. See \cite{Sperner1928}.}
	Indeed, toward a contradiction, assume there are $q_1,q_2 \in Q$ such that $I(q_1)\subseteq I(q_2)$.
	We obtain that $\bigcap_{i \in I(q_1)} S_i \supseteq \{q_1,q_2\}$, a contradiction.
	By Sperner's theorem, we get $\vert Q \vert \leq$ $n \choose \lfloor \frac{n}{2} \rfloor$.
	Hence if $\{S_1, \ldots , S_n\}$ is a selective family, then $\vert Q \vert \leq$ $n \choose \lfloor \frac{n}{2} \rfloor$.
    
    Next we show that for each $n$ such that $\vert Q \vert \leq$ $n \choose \lfloor \frac{n}{2} \rfloor$, there is a selective family of size $n$ and conclude that $\mathrm{sel}(Q)$ is the minimum of such $n$.
    Let $n$ be such that $\vert Q \vert \leq$ $n \choose \lfloor \frac{n}{2} \rfloor$ and define an injective function $f$ from $Q$ to $\lfloor \frac{n}{2} \rfloor$-subsets of $\{1,\ldots, n\}$ (this can always be done, since $n \choose \lfloor \frac{n}{2} \rfloor$ is the number of subsets of size $\lfloor \frac{n}{2} \rfloor$ in a set of size $n$).
	Since for all $q \in Q$ the subsets $f(q)$ are of same cardinality and are distinct, no $f(q)$ is contained in another one. 
	For all $p=1,\ldots, n$ set $S_p:=\{q: p \in f(q)\}$, we show that $\{q\}=\bigcap_{p \in f(q)}S_p$ for any $q\in Q$.
	The definition of each $S_p$ implies that $q \in \bigcap_{p \in f(q)}S_p$.
	For the inclusion, toward a contradiction assume there is $q^{\prime} \neq q$ such that $q^{\prime} \in \bigcap_{p \in f(q)}S_p$.
	We obtain that for all $p \in f(q)$, $q^{\prime} \in S_p$ holds.
	Hence, by definition of $S_p$, we get that if $p \in f(q)$, then $p \in f(q^{\prime})$.
	We conclude that $f(q) \subseteq f(q^{\prime})$.
	Since $\vert f(q) \vert = \vert f(q^{\prime}) \vert$, we have $f(q)=f(q^{\prime})$, a contradiction to $f$ being injective.
\end{proof}

For instance, suppose we have ten relevant dimension in a joint choice problem, that is, $\vert Q \vert = 10$. 
The least integer $n$ such that $\vert Q\vert\leq $$n \choose \lfloor \frac{n}{2} \rfloor$ is $5$.
By Lemma~\ref{LEMMA: smallest Martin family}, to check separability on a complete joint choice on ten dimensions, it is enough to test $S_i$-separability for any $S_i$ in a selective family $\mathscr {S}$ of size five.
 

\section{Rationalizability and separability} \label{SECT:rational_vs_separable}

To conclude our analysis, we relate separability with a classical property of choices, namely rationalizability.
Moreover, we investigate the relation between separable joint choices, separable preferences, and additively separable utility functions. 

To start, we adapt some known notions to our new setting. 
A \textit{joint preference} on $\prod_{q \in Q} X_q$ is a reflexive binary relation $\succsim$ on $\prod_{q \in Q} X_q$ such that its asymmetric part $\succ$ is \textit{acyclic}, that is, there are no $x_Q^1,x_Q^2,\ldots,x_Q^k \in \prod_{q \in Q} X_q$ (with $k \geqslant 3$) such that $x_Q^1 \succ x_Q^2 \succ \ldots x_Q^k \succ x_Q^1$.\footnote{Note that $\succsim$ is the union of its asymmetric part $\succ$ and its symmetric part $\sim$. As usual, $\succ$ is defined by $x_Q \succ y_Q$ if $x_Q \succsim y_Q$ and $\neg(y_Q \succsim x_Q)$, whereas $\sim$ is defined by $x_Q \sim y_Q$ if $x_Q \succsim y_Q$ and $y_Q \succsim x_Q$.} 
The set of \textit{$\succsim$-maximal} elements of a multidimensional menu $A_{Q} \subseteq \prod_{q \in Q} X_q$ is then defined by 
$$
\max(A_Q, \succsim) = \{x_Q \in A_Q \colon y_Q \succ x_Q \text{ for no } y_Q \in A_Q\}.
$$
Note that $\max(A_Q, \succsim)$ is nonempty by the finiteness of $\prod_{q \in Q} X_q$ 
	and the acyclicity of $\succsim$. 

We now define the twin notions of `revealed joint choice' and `rationalizable joint choice'.
{
As for the classical case,\footnote{Rationalizability was originally defined first for choice functions by \cite{Samuelson1938} and \cite{Houthakker1950}, and then for choice correspondences by \cite{Arrow1959}, \cite{Ricther1966} and \cite{Sen1971}, among others. }  multidimensional rationalizability requires that the alternatives selected in each menu are maximal with respect to the preference revealed from the observed joint choice. 
}

\begin{definition}\label{DEF:revealed_joint_choice}
Given a joint preference $\succsim$ on $\prod_{q \in Q} X_q$, the \textit{revealed joint choice}  $c^{\succsim}$ on $\mathfrak{M}_{Q}$ is the map
$$
c^{\succsim}\colon \mathfrak{M}_{Q}\to 2^{\prod_{q \in Q} X_q}
\,,\quad  c^{\succsim}(A_Q)=\max(A_Q,\succsim)\quad \text{for all } A_Q\in \mathfrak{M}_{Q}.
$$ 
Conversely, given a joint choice $c \colon \mathfrak{M}_{Q} \to 2^{\prod_{q \in Q} X_q}$ on $\mathfrak{M}_{Q}\subseteq\prod_{q\in Q}2^{X_q}$, the \textit{revealed joint preference} is the binary relation $\succsim^c$ on $\prod_{q \in Q} X_q$ defined as follows for all $x_Q,y_Q \in \prod_{q \in Q} X_q$:
$$
x_Q \succsim^c y_Q \quad \stackrel{\mathrm{def}}{\Longleftrightarrow} \quad 
\text{there is }\; B_Q \in \mathfrak{M}_{Q} \;\;\text{ such that }\;\; x_Q \in c(B_Q) \text{ and } y_Q \in B_Q\,.
$$
Then $c$ is \textit{rationalizable} if $c(A_Q) = \max(A_Q,\succsim^c)=c^{\succsim^{c}}(A_Q)$ for any menu $A_Q \in \mathfrak{M}_{Q}$.\footnote{Note that the terminology is sound, because $\succsim^c$ is a joint preference on $\prod_{q \in Q} X_q$.} 
\end{definition}

Rationalizability and separability are independent notions, as the next counterexample shows. 

\begin{counterexample}[\it Independence of separability and rationalizability] \label{EX:rationalizability_and_separability_are_independent}
	We exhibit two complete joint choices: the first is rationalizable but not separable, whereas the second is separable but not rationalizable. 
	Both joint choices are $2$-dimensional, that is, $Q=\{1,2\}$, with $X_1=ab$ and $X_2=xy$. 
	Let $c\colon 2^{X_1}\times 2^{X_2}\to 2^{X_1\times X_2}$ be defined by
	\begin{center}
		\smallskip
		\begin{tabular}
		{|c||c|c|c|}
		\hline
		$c(\cdot,\cdot)$ & $xy$ & $x$ & $y$\\
		\hline\hline
		$ab$ & $\{(a,x),(b,y)\}$ & $\{(a,x)\}$ & $\{(b,y)\}$ \\
		\hline
		$a$ & $\{(a,x)\}$ & $\{(a,x)\}$ & $\{(a,y)\}$\\
		\hline
		$b$ & $\{(b,y)\}$ & $\{(b,x)\}$ & $\{(b,y)\}$ \\
		\hline
		\end{tabular}
	\smallskip
	\end{center}
	 It is immediate to check that $c$ is not $\{1\}$-separable, hence it is not separable by Theorem~\ref{THM:q_separability_equivalent_separability_by_dimensions}.
	 However, $c$ is rationalizable.
	 In fact, the revealed preference $\succsim^{c}$, which is defined by
	 $$
	 (a,x)\sim^{c} (b,y)\succ^{c} (b,x)\,,\qquad (a,x)\succ^{c} (b,x)\,,\qquad (a,x)\succ^{c}(a,y)\,,\qquad (b,y)\succ^{c}(a,y)\,
	 $$  
	 is such that $c(A_{1,2})=\max(A_{1,2},\succsim^c)$ for any $A_{1,2} \in 2^{X_1} \times 2^{X_2}$. 
 \smallskip
 
 Consider now the joint choice $c^{\prime}\colon 2^{X_1}\times 2^{X_2}\to 2^{X_1\times X_2}$ defined by 	
	\begin{center}
		\smallskip
		\begin{tabular}	{|c||c|c|c|}
		\hline
		$c^{\prime}(\cdot,\cdot)$ & $xy$ & $x$ & $y$\\
		\hline\hline
		$ab$ & $\{(a,x),(b,y)\}$ & $\{(a,x),(b,x)\}$ & $\{(a,y),(b,y)\}$ \\
		\hline
		$a$ & $\{(a,x),(a,y)\}$ & $\{(a,x)\}$ & $\{(a,y)\}$\\
		\hline
		$b$ & $\{(b,x),(b,y)\}$ & $\{(b,x)\}$ & $\{(b,y)\}$ \\
		\hline		
		\end{tabular}
	\smallskip
	\end{center}
Since the equalities
\begin{itemize}
	\item $\pi_1(c^{\prime}(ab,xy))=\pi_1(c^{\prime}(ab,x))=\pi_1(ab,y)=ab$,
	\item  $\pi_1(c^{\prime}(a,xy))=\pi_1(c^{\prime}(a,x))=\pi_1(c^{\prime}(a,y))=a$,
	\item $\pi_1(c^{\prime}(b,xy))=\pi_1(c^{\prime}(b,x))=\pi_1(c^{\prime}(b,y))=b$
\end{itemize}
	 hold, $c^{\prime}$ is $\{1\}$-separable.
	 Moreover, since the equalities
	 \begin{itemize}
	 	\item $\pi_2(c^{\prime}(ab,xy))=\pi_2(c^{\prime}(a,xy))=\pi_2(b,xy)=xy$,
	 	\item $\pi_2(c^{\prime}(ab,x))=\pi_2(c^{\prime}(a,x))=\pi_2(b,x)=x$,
	 	\item $\pi_2(c^{\prime}(ab,y))=\pi_2(c^{\prime}(a,y))=\pi_2(b,y)=y$,
	 \end{itemize}
	 hold, we deduce that $c^{\prime}$ is also $\{2\}$-separable.
	 Thus, $c^{\,\prime}$ is separable by Theorem~\ref{THM:q_separability_equivalent_separability_by_dimensions}.  
	 On the other hand, $c^{\,\prime}$ is not rationalizable.
	 In fact, the revealed preference $\succsim^{c^{\prime}}$, which is defined by 
	 $$ 
	 (a,x)\sim^{c^{\prime}} (b,x) \sim^{c^{\prime}} (b,y) \sim^{c^{\prime}}(a,x) \sim^{c^{\prime}} (a,y) \sim^{c^{\prime}} (b,y),
	 $$
	 
	 is such that $\max((ab,xy),\succsim^{c^{\prime}})=\{(a,x),(a,y),(b,y)\}$, but $c^{\prime}(ab,xy)=\{(a,x),(b,y)\}$. 
\end{counterexample}

Although separability and rationalizability are well distinct properties, their interplay determines the relation that connects separable joint choices and separable joint preferences, which are defined as follows.

\begin{definition}[\citealp{BradleyHodgeKilgour2005}] \label{DEF:separable_preferences} 
	For any nonempty $S \subseteq Q$, a joint preference $\succsim$ on $\prod_{q \in Q} X_q$ is \textit{$S$-separable} if the implication 
	\begin{align*}
	(x_S, u_{-S}) \succsim (y_S, u_{-S}) \,\text{ for some }\, u_{-S} \in \!\prod_{q \in -S} X_q & \ \Longrightarrow \
	(x_S, u_{-S}) \succsim (y_S, u_{-S}) \,\text{ for all }\, u_{-S} \in \!\prod_{q \in -S} X_q
	\end{align*}
	holds for all $x_S, y_S \in \prod_{q \in S} X_q$. 
	Then $\succsim$ is \textit{separable} if it is $S$-separable for all nonempty $S \subseteq Q$. 
	If $\succsim$ is a $S$-separable joint preference, then the joint preference \textit{induced on $S$} is the binary relation $\succsim_S$ on $\prod_{q \in S} X_q$ defined as follows for any $x_S,y_S \in \prod_{q \in S} X_q$: 
	$$
	x_S \succsim_S y_S \quad \Longleftrightarrow \quad (x_S, u_{-S}) \succsim (y_S, u_{-S}) \;\text{ for some (equivalently, for all) }\; u_{-S} \in \prod_{q \in -S} X_q.
	$$ 
\end{definition}

As for joint choices, $S$-separability of discrete preferences posits the existence of a hidden consistent preference over alternatives with dimensions in $S$.
Since Definition~\ref{DEF:separable_preferences} has the same rationale as the characterization of separability of joint choices given in Lemma~\ref{LEMMA:chz_for_separability}, one may wonder what is the link between these two notions.  
It turns out that separability of joint choices extends separability of joint preferences.
To see why, we first have to introduce a property of the choice domain.




\begin{definition}\label{DEF:menus_S_composition}
	Fix a nonempty set $S\subseteq Q$ of dimensions. 
	A family $\mathfrak{M}_{Q}\subseteq  \prod_{q \in Q} 2^{X_q}$ of menus is \textit{$S$-rich} if the existence of some $A_Q \in \mathfrak{M}_{Q}$ such that $(x_S, u_{-S}), (y_S, u_{-S}) \in A_Q$ implies that 
	$$
	\Bigg(\prod_{q \in S} \big(\pi_q(x_S) \cup \pi_q(y_S)\big), v_{-S}\Bigg) \in \mathfrak{M}_{Q},
	$$
	for all $v_{-s} \in \prod_{q \in -S} X_q$.\footnote{Here we slightly abuse notation: $\Big(\prod_{q \in S} \big(\pi_q(x_S) \cup \pi_q(y_S)\big), v_{-S}\Big)$ is to be interpreted as the menu composed of binary sets $\{\pi_q(x_S), \pi_q(y_S)\}$ in dimensions $q \in S$ and singletons $\{v_q\}$ in dimensions $q \in -S$. 
	}
\end{definition}


 
$S$-richness ensures the existence of minimal menus in which $x_S$ and $y_S$ are available, combined to any possible $u_{-S}$ in the remaining dimensions. 
%
	For instance, assume that $Q=\{1,2,3\},$ $X_1=ab,$ $X_2=d,$ $X_3=xy.$ 
	Suppose also that $(ab,d,x)\in \mathfrak{M}_{Q},$ and consider the subset $\{1,2\}$ of dimensions.
Notice that $(a,d,x)$ and $(b,d,x)$ belong to $(ab,d,x).$
Thus, $\{1,2\}$-richness requires that $(ab,d,y)\in\mathfrak{M}_Q$.\\
 
The property of $S$-richness is rather strong, because it assumes that any menu obtained by joining two available alternatives in each dimension $q\in S$ is feasible; however, it is needed to retrieve a $S$-separable preference from a $S$-separable joint choice.    
In fact, whenever we witness a preference between $(x_S,u_{-S}),(y_S,u_{-S})$, $S$-richness always allows to find a menu to check the preference is consistent for any other $v_{-S}$ we may couple to $x_S$ and $y_S$.


\begin{theorem}\label{THM:separabilty_preference_vs_choice}
Let $S \subseteq Q$ and $\mathfrak{M}_{Q}\subseteq\prod_{q\in Q}2^{X_q}$.
\begin{itemize}
	\item[\rm (i)] If a joint preference $\succsim$ on $\prod_{q \in Q} X_q$ is $S$-separable, then the revealed joint choice $c^{\succsim}$ on $\mathfrak{M}_{Q}$ is $S$-separable. 
	\item[\rm (ii)] If a joint choice $c$ on $\mathfrak{M}_{Q}$ is both $S$-separable and rationalizable, and $\mathfrak{M}_Q$ is \textit{$S$-rich}, then the revealed joint preference $\succsim^{c}$ on $\prod_{q \in Q} X_q$ is $S$-separable.  
\end{itemize}
\end{theorem}

\begin{proof}
	\underline{(i):} Let $\succsim$ be a  $S$-separable joint preference on $\prod_{q \in Q} X_q$.   
	Consider two menus $(A_S,B_{-S})$ and $(A_S,C_{-S})$ belonging to $\in\mathfrak{M}_Q$, where $A_{S}\in\pi_S(\mathfrak{M}_Q),$ and $B_{-S},C_{-S}\in\pi_{-S}(\mathfrak{M}_Q).$
	The definition of $c^{\succsim}$ yields $\pi_S(c^{\succsim}(A_S,B_{-S}))=\pi_{S}(\max((A_S,B_{-S}),\succsim))$ and $\pi_S(c^{\succsim}(A_S,C_{-S}))=\pi_{S}(\max((A_S,C_{-S}),\succsim))$.
	Moreover, $S$-separability of $\succsim$ implies $\pi_{S}(\max((A_S,B_{-S}),\succsim))=\pi_{S}(\max((A_S,C_{-S}),\succsim)$. 
	We get $\pi_S(c^{\succsim}(A_S,B_{-S}))=\pi_S(c^{\succsim}(A_S,C_{-S}))$, which by Lemma~\ref{LEMMA:chz_for_separability} implies that $c^{\succsim}$ is $S$-separable. 
\smallskip
	
		\underline{(ii):}
	 Let $c\colon \mathfrak{M}_{Q} \to 2^{\prod_{q\in Q}\X_q}$ be a $S$-separable and rationalizable joint choice function on $\mathfrak{M}_Q$, and assume that $\mathfrak{M}_Q$ is $S$-rich.
	  Consider alternatives $(x_S,u_{-S}),(y_S,u_{-S})\in \prod_{q \in Q} X_q$, with $x_S,y_S\in \prod_{q \in S} X_q$, and $u_{-S} \in \prod_{q \in -S}{X_q}$, and assume that $(x_S, u_{-S}) \succsim^{c} (y_S, u_{-S})$.
	The definition of $\succsim^{c}$ implies that there is a menu $A_Q \in \mathfrak{M}_Q$ such that $(x_S, u_{-S}), (y_S, u_{-S}) \in A_Q$ and $(x_S, u_{-S})\in c(A_Q)$. 
	Since $\mathfrak{M}_Q$ is $S$-rich and $c$ is rationalizable, we have $(x_S, u_{-S}) \in c\left(\prod_{q \in S} \big(\pi_q(x_S) \cup \pi_q(y_S)\big), u_{-S}\right)$.
	$S$-separability of $c$ implies that $\pi_{S}\left(c\left(\prod_{q \in S} \big(\pi_q(x_S) \cup \pi_q(y_S)\big), u_{-S}\right)\right) = c_{S}\left(\prod_{q \in S} \big(\pi_q(x_S) \cup \pi_q(y_S)\big)\right)$, with $x_S$ belonging to $c_{S}\left(\prod_{q \in S} \big(\pi_q(x_S) \cup \pi_q(y_S)\big)\right)$.
	Consider a menu $\left(\prod_{q \in S} \big(\pi_q(x_S) \cup \pi_q(y_S)\big), v_{-S}\right)$, with $v_{-S}\in \prod_{q \in -S} X_q$. 
	Since $\mathfrak{M}_Q$ is $S$-rich, $\left(\prod_{q \in S} \big(\pi_q(x_S) \cup \pi_q(y_S)\big), v_{-S}\right)\in \mathfrak{M}_Q.$
	$S$-separability yields $\pi_{S}\left(c\left(\prod_{q \in S} \big(\pi_q(x_S) \cup \pi_q(y_S)\big), v_{-S}\right)\right) = c_{S}\left(\prod_{q \in S} \big(\pi_q(x_S) \cup \pi_q(y_S)\big)\right)$. 
	This implies that $(x_S, v_{-S}) \in c\left(\prod_{q \in S} \big(\pi_q(x_S) \cup \pi_q(y_S)\big), v_{-S}\right)$.
	The definition of $\succsim^{c}$ yields $(x_S,v_{-S})\succsim^{c}(x_S,y_{-S}),$ which concludes the proof.
%
\end{proof}

Thus, $S$-separable joint preferences determine $S$-separable joint choices, but the converse is not always true.
Indeed, only rationalizable and $S$-separable joint choices defined on a domain that is $S$-rich reveal $S$-separable joint preferences on the alternatives.
\footnote{{\cite{BradleyHodgeKilgour2005} show that $S$-separability of preferences is not stable under union, that is, $\succsim$ is $S$-separability and $T$-separability of $\succsim$ does not imply $(S \cup T)$-separability. On the contrary, we have proved that $S$-separability if stable under union for choice. This may seem at odds with the result in Theorem~\ref{THM:separabilty_preference_vs_choice}. Yet, the conflict is easily resolved upon noting that rationalizable choices reveal incomplete preferences, by definition of the domain, which necessarily exclude pairs contrasting with stability under union.
}}

The proof of Theorem~\ref{THM:separabilty_preference_vs_choice} relies on two observations.
First, the complexity of joint choices cannot be always captured by the binariness of joint preferences.  
Rationalizability allows to collapse a multidimensional choice to a multidimensional preference, and transfers the $S$-separability of the choice onto the preference. 
Second, a $S$-separable preference, once the projections over $S$ have been fixed, compares all alternatives which differ in the dimensions belonging to $-S$. 
This amount of information can be certainly retrieved from a joint choice if the choice domain is $S$-rich. 
When one of the two properties is not satisfied, $S$-separable joint choices need not induce $S$-separable joint preferences, as the next counterexample shows. 

\begin{counterexample}[\it Separability and rationalizability for joint choices and joint preferences]\label{EXM:rationalizable_not_menus_S_combination_and_viceversa}
	We exhibit two joint choices on $Q=\{1,2\}$, with $X_1=ab$ and $X_2=xy$. 
	 The first choice is $S$-separable for some $S\subseteq Q$ and satisfies $S$-richness, but fails to be rationalizable. 
	  The second choice is rationalizable and $S$-separable for some $S\subseteq Q$, but is not $S$-rich.
	  We show that in both cases the revealed joint preference is not $S$-separable.

	Let $c\colon 2^{X_1}\times 2^{X_2}\to 2^{X_1\times X_2}$ be the complete joint choice defined by
	\begin{center}
		\smallskip
		\begin{tabular}{|c||c|c|c|}
		\hline
		$c(\cdot,\cdot)$ & $xy$ & $x$ & $y$\\
		\hline\hline
		$ab$ & $\{(a,x)\}$ & $\{(a,x)\}$ & $\{(a,y)\}$ \\
		\hline
		$a$ & $\{(a,x),(a,y)\}$ & $\{(a,x)\}$ & $\{(a,y)\}$\\
		\hline
		$b$ & $\{(b,y)\}$ & $\{(b,x)\}$ & $\{(b,y)\}$ \\
		\hline	
		\end{tabular}
	\smallskip
	\end{center}
To see that $c$ is $\{1\}$-separable, observe that $\{a\}=\pi_1(c(ab,xy))=\pi_1(c(ab,x))=\pi_1(c(ab,y))$.
Moreover, $c$ is complete, which implies that $\mathfrak{M}_{\{1,2\}}=2^{X_1}\times 2^{X_2}$ is $\{1\}$-rich.
However, $c$ is not rationalizable.
Indeed, the revealed preference $\succsim^{c}$ is defined by 
$$
(a,x)\sim^{c}(a,y)\,,\qquad (a,x)\succ^{c}(b,x)\,,\qquad(a,x)	\succ^{c}(b,y)\,,\qquad (a,y)	\succ^{c}(b,y)\,,\qquad (b,y)\succ^{c}(b,x)\,,
$$
but $\max((ab,xy),\succsim^{c})=\{(a,x),(a,y)\} \neq \{(a,x)\}=c(ab,xy)$.

Consider now the joint choice $c^{\prime}\colon \mathfrak{M}_{1,2} \to 2^{X_1\times X_2}$ on the domain $\mathfrak{M}_{1,2} = \{(ab,xy),(ab,x)\}$ defined by 
	\begin{center}
	\smallskip
	\begin{tabular}{|c||c|c|}
		\hline
		$c^{\prime}(\cdot,\cdot)$ & $xy$ & $x$ \\
		\hline\hline
		$ab$ & $\{(a,x), (b, y)\}$ & $\{(a,x)\}$ \\
		\hline	
	\end{tabular}
	\smallskip
\end{center}
Then the revealed preference $\succsim^{c^{\prime}}$ is given by
$$
	(a,x)\sim^{c^{\prime}} (b,y)\succ^{c^{\prime}}(b,x)\,,\qquad  (a,x)\succ^{c^{\prime}}(a,y)\,,\qquad (a,x) \succ^{c^{\prime}}(b,x)\,,\qquad (b,y)\succ^{c^{\prime}}(a,y)\,.
$$
Since $c^{\prime}(ab,xy)=\{(a,x),(b,y)\}=\max((ab,xy),\succsim^{c^{\prime}})$ and  $c^{\prime}(ab,x)=\{(a,x)\}=\max((ab,x),\succsim^{c^{\prime}})$ hold, we conclude that $c^{\prime}$ is rationalizable. 
Moreover, 
$c^{\prime}$ is also $\{2\}$-separable. 
However, $c^{\prime}$ is not $\{2\}$-rich, because neither $(a,xy)$ nor $(b,xy)$  belong to $\mathfrak{M}_{1,2}$.  
To see that $\succsim^{c^{\prime}}$ is not $\{2\}$-separable, note that $(a,x)\succ^{c^{\prime}}(a,y)$ and $(b,y)\succ^{c^{\prime}}(b,x)$ hold.
\end{counterexample}

Separable joint choices extend also a class of separable utility functions. 
	Separable utilities are usually defined on an infinite domain, and the investigation of their formal and conceptual features requires differentiability.\footnote{In a constrained optimization problem with positive prices and finite income with a $S$-separable utility function, the optimal allocation of expenditure within set $S$ preserves relative proportions in response to changes in income or prices pertaining to other dimensions.} 
	For the scope of this note, we rather focus on separability as a property induced by the functional form of utility. 
	Following \cite{Strotz1957} and \cite{Gorman1959}, a separable utility function is of the kind
	$$
	U = U[V_R(x_R), V_S(x_S), \dots, V_T(x_T)]
	$$
	where 
	\begin{itemize}
		\item $\{R, S, \dots, T\}$ is a partition of the set of dimensions $Q$,
		\item $x_R \in \prod_{q \in R}X_q$, $x_S \in \prod_{q \in S}X_q$, \dots, $x_T \in \prod_{q \in T}X_q$ are partial alternatives, 
		\item $V_R$, $V_S$, \dots, $V_T$ are \textit{branch} utility functions. 
	\end{itemize} 
	Most commonly, separable utility functions are required to be additive: 
	$$
	U = V_R(x_R) + V_S(x_S) + \dots + V_T(x_T).
	$$
	In our setting, any additively separable utility function where dimensions in $S$ are aggregated induces a $S$-separable joint choice.
	
	\begin{lemma} \label{LEMMA:additive utility}
		Let $U \colon \prod_{q \in Q}X_q \to \mathbb{R}$ be such that 
		$$
		U(x_Q)= V_R(x_R) + V_S(x_S) + \dots + V_T(x_T)\,,
		$$ 
		with $V_R \colon \prod_{q \in R}X_q \to \mathbb{R}$, $V_S \colon \prod_{q \in S}X_q \to \mathbb{R}$, \dots, $V_T \colon \prod_{q \in T}X_q \to \mathbb{R}$. 
		Then any joint choice $c \colon 	\mathfrak{M}_{Q} \to 2^{\prod_{q \in Q} X_q}$  on $\mathfrak{M}_{Q}\subseteq \prod_{q\in Q} 2^{X_q}$ such that $c(A_Q) = \arg \max_{x_Q \in A_Q} U(x_Q)$ is $S$-separable. 
	\end{lemma} 

	\begin{proof}
		Note that 
		\begin{align*}
			U(c(A_Q)) &=	\max_{x_Q \in A_Q} U(x_Q) \\
				& = \max_{x_Q \in A_Q} V_R(\pi_{R}(x_Q)) + V_S(\pi_{S}(x_Q)) + \dots + V_T(\pi_{T}(x_Q)) \\
				&= \max_{x_R \in A_R} V_R(x_R) + \max_{x_S \in A_S} V_S(x_{S}) + \dots + \max_{x_T \in A_T}V_T(x_{T}).
		\end{align*}
		It follows that 
		$$
		\pi_S(c(A_S,B_{-S}))= \arg\max_{x_S \in A_S}  V_S(x_{S}) = \pi_S(c(A_S,C_{-S}))
		$$ 
		for all $A_S \in \pi_S(\mathfrak{M}_Q)$ and $B_{-S}, C_{-S} \in \pi_{-S}(\mathfrak{M}_Q)$.
	\end{proof}

	This simple lemma complements a known result in \cite{BradleyHodgeKilgour2005}.


\begin{definition}[\citealp{BradleyHodgeKilgour2005}] 
	Let $\succsim$ be a total preorder\footnote{A \textit{total preorder} is a reflexive, transitive, and complete binary relation.} on $\prod_{q \in Q} X_q$. 
	A \textit{value function} for $\succsim$ is a map $v\colon \prod_{q \in Q} X_q\to \mathbb{R} $ such that for any $x_Q,y_Q\in \prod_{q \in Q} X_q$, we have $x_Q\succsim y_Q$ if and only if $v(x_Q)\succsim v(y_Q)$. 
	Moreover, $\succsim$ is \textit{additive} if for any $q\in Q$, there is a map  $v_q\colon X_q \to \mathbb{R}$ such that $v(x_Q)=\sum_{q\in Q} v_q(x_q)$ for any $x_Q\in \prod_{q \in Q} X_q$; in this case, each $v_q$ is called a \textit{criterion value}. 
\end{definition}  

An additive joint preference is represented by an additive separable utility function obtained by summing $\vert Q\vert$ mono-dimensional utilities.
Additive preferences are also separable:

\begin{lemma}[\citealp{Yu1985,BradleyHodgeKilgour2005}]
	For any joint preference $\succsim$, additivity implies separability. 
\end{lemma} 

Using Theorem~\ref{THM:separabilty_preference_vs_choice}, we get the following result.

\begin{corollary}\label{COR:additive_preference_induces_separable_choice}
	For any joint preference $\succsim$, if $\succsim$ is additive, then the revealed joint choice $c^{\succsim}$ is separable.
\end{corollary}

The converse of Lemma~\ref{LEMMA:additive utility} and Corollary~\ref{COR:additive_preference_induces_separable_choice} do not hold.
Indeed, the joint choice $c^{\prime}$ exhibited in Example~\ref{EX:rationalizability_and_separability_are_independent} is separable but not rationalizable, and so Theorem~\ref{THM:separabilty_preference_vs_choice} and Corollary~\ref{COR:additive_preference_induces_separable_choice} imply that its revealed joint preference is not additive.
 Since $c^{\prime}$ is defined for two dimensions only, there is also no additive separable utility function which recovers $c^{\prime}$.

 We conclude this section with a remark that enables the experimenter to test rationalizability of the full dataset by verifying rationalizability of choices defined on fewer dimensions.
Specifically, it shows that rationalizability of a separable joint choice can be elicited by rationalizability of the joint choices induced by the subsets of dimensions of \textit{any} selective family.

\begin{theorem} \label{THM: local rat is rat}
	Let $c \colon \mathfrak{M}_{Q} \to 2^{\prod_{q \in Q} X_q}$ be a separable joint choice.
	The following statements are equivalent:
	\begin{itemize}
		\item [\rm(i)] $c$ is rationalizable;
		\item [\rm(ii)] for any selective family $\mathscr S$ and any $S \in \mathscr S$, the induced joint choice $c_S$ is rationalizable;
		\item [\rm(iii)] there is a selective family $\mathscr S$ such that for each $S \in \mathscr S$, the induced joint choice $c_S$ is rationalizable.
	\end{itemize}
\end{theorem}

\begin{proof}
	\underline{(i)$\implies$(ii):} 
	Suppose $c$ is rationalizable, and let $\mathscr S$ be a selective family for $Q$. 
	Take any $S \in \mathscr S$, and consider the induced joint choice $c_S \colon \pi_{S}(\mathfrak{M}_Q) \to 2^{\prod_{q \in S} X_q}$.
	The definition of $c_S$ implies that $c_S(A_S)=c_S(\pi_{S}(A_Q))=\pi_{S}(c(A_{Q}))$ for any $A_Q \in \mathfrak{M}_Q$.
	 Rationalizability of $c$ yields $c(A_Q) = \max(A_Q,\succsim^c)$ for any $A_Q \in \mathfrak{M}_Q$.
	By Theorem~\ref{THM:separabilty_preference_vs_choice}, the revelaled joint preference $\succsim^{c}$ is $S$-separable, which implies that $\pi_S(\max(A_Q,\succsim^c))=\max(A_S,\succsim^{c_S})$ for any $A_Q \in \mathfrak{M}_Q$. 
	Since $c_S(A_S)=c_S(\pi_{S}(A_Q))=\pi_{S}(c(A_{Q}))=\pi_{S}(\max(A,\succsim^c))=\max(A_S,\succsim^{c_S})$ for any $A_Q \in \mathfrak{M}_Q$, we conclude that $c_S$ is rationalizable.
	\smallskip
	
	\underline{(ii)$\implies$(iii):}
	Obvious.
	\smallskip
	
	\underline{(iii)$\implies$(i):}
	Let $\mathscr S$ be a selective family for $Q$ such that for any $S \in \mathscr S$,  the induced joint choice $c_S \colon \pi_{S}(\mathfrak{M}_Q) \to 2^{\prod_{q \in S} X_q}$ is rationalizable.
	For each $S \in \mathscr S$, consider the family $\mathscr{S}_{S}=\{\{a\}:a \in S\}$ of singletons, which is obviously selective for $S$. 
	Since (i) implies (ii), 
	the joint choice ${c_{{S}_{\{_a\}}}}={c_{{S}_{_a}}}$ induced by $\{a\}$ is rationalizable.
	Note that ${c_{{S}_{_a}}}=c_{a}$ for all $S \in \mathscr S$ and $a \in S$.
	Thus, $c_{a}$ is rationalizable for all $a\in S$ and $S \in \mathscr S$.
	Since $\mathscr S$ is a selective family for $Q$, we obtain that $c_{q}$ is rationalizable for each $q \in Q$.
	To complete the proof, we now show that $c$ is rationalizable.

The definition of $\succsim^{c}$ implies that $c(A_{Q})\subseteq \max (A_{Q},\succsim^{c})$ for any $A_{Q}\in \mathfrak{M}_Q$.
Thus, it suffices to prove hat $c(A_{Q})\supseteq \max(A_{Q},\succsim^{c})$ for any $A_{Q}\in \mathfrak{M}_Q$.
Toward a contradiction, assume that there are $A_{Q}\in \mathfrak{M}_Q$ and $x_{Q}\in A_{Q}$ such that $x_{Q}\in\max(A_{Q},\succsim^{c})$ but $x_{Q}\not\in c(A_Q)$. 
The definition of $c_{q}$ implies that there is $q\in Q$ such that $\pi_{q}(x_{Q})\not\in c_{q}(A_Q)$. 
Since $c_{q}$ is rationalizable for each $q \in Q$ (as proved in the previous paragraph), we conclude that $\pi_{q}(x_{Q}) \notin \max(\pi_{q}(A_{Q}),\succsim^{c_{q}}).$ 
Thus, we obtain that  $x_{Q}\in\max(A_{Q},\succsim^{c})$ and $\pi_{q}(x_{Q})\not\in\max(\pi_{q}(A_{Q}),\succsim^{c_{q}})$ for some $q\in Q$.
This is impossible, since the definition of revealed joint preference implies that, for all $x_Q,  y_Q \in \prod_{q \in Q} X_q$ and $q \in Q$,  if $x_Q \succsim^c y_Q $, then $x_q \succsim^{c_{q}} y_q$. 
\end{proof}

\section{Conclusions} \label{SECT:conclusions}

In this work we have introduced joint choices, which display multiple selections from menus containing multidimensional alternatives.
$S$-separability has been defined for joint choices, imposing that picks associated to dimensions in $S$ are independent from the remaining dimensions. 
Separability of joint choices requires $S$-separability of the dataset for any subset $S$ of dimensions.  
Stability of $S$-separability has been thoroughly investigated, showing that it is preserved under unions and --- whenever the choice domain satisfies a closure property --- under intersections. 
Thanks to these properties, an experimenter can significantly reduce the verification steps to elicit separability.
Upon defining rationalizability of joint choices, we have shown that separability of joint choices extends separability of discrete preferences and utility functions.
Finally, when separability holds, rationalizability can be characterized by the rationalizability of joint choices induced by subsets of dimensions belonging to some selective family.

Our theoretical analysis can be continued in at least two directions. 
First, an empirical or experimental investigation of separable joint choices would further validate our work.
In this direction of research, the main difficulty is the construction of a dataset that respects some of the properties listed in Definitions~\ref{DEF:menus_betweenness}, \ref{DEF:chained_menus_betweennes}, and \ref{DEF:menus_S_composition}.
 For instance, a real-world joint choice whose domain satisfies menus betweenness with respect to some pair of dimensions should gather selections from several multidimensional menus. 
With a high number of dimensions, such a task may appear problematic. 
However, many results achieved in this work do not rely on any feature of the choice domain.  
Moreover, when few dimensions are considered, joint choices defined on a suitably large amount of menus can be more easily obtained from lab experiments  and empirical observations, such as the consumption scanner data adopted in \cite{deHaanvanderGrient2011} and  \cite{RenkinMontialouxSiegenthaler2022}.

A different --- and possibly more interesting --- direction of research is that of the analysis of the separability of \textit{stochastic joint choices}, which report, for any menu, selection probabilities of each feasible subset of multidimensional alternatives.
As already mentioned in the Introduction, the basis for this investigation should be the work of \cite{ChambersMasatliogluTuransick2024} and \cite{KashaevPlavalaAguiar2024}, who have already studied separable two-dimensional stochastic joint choice functions. 



\end{document}